\documentclass[aps,amsmath,amssymb]{revtex4}
\usepackage{graphicx}

\usepackage{amsthm}

\usepackage{bm}
\usepackage{float}
\usepackage{mathtools}
\usepackage{multirow}
\usepackage{commath}
\usepackage{braket}
\usepackage{subfig}
\usepackage{bbold}
\usepackage{mathrsfs}
\usepackage[usenames]{xcolor}
\usepackage{gensymb}
\usepackage{appendix}
\usepackage{graphicx}
\usepackage{subfig}

\usepackage[normalem]{ulem}

\DeclareMathOperator{\Tr}{Tr}

\begin{document}
\def\mean#1{\left< #1 \right>}

\title{Efficiency fluctuations in a quantum battery charged by a repeated interaction process}

\author{Felipe Barra\footnote{fbarra@dfi.uchile.cl}}
\affiliation{Departamento de F\'isica, Facultad de Ciencias F\'isicas y Matem\'aticas, Universidad de Chile, Santiago, Chile}

\begin{abstract}

A repeated interaction process assisted by auxiliary thermal systems charges a quantum battery. The charging energy is supplied by switching on and off the interaction between the battery and the thermal systems. The charged state is an equilibrium state for the repeated interaction process, and the ergotropy characterizes its charge. The working cycle consists in extracting the ergotropy and charging the battery again. We discuss the fluctuating efficiency of the process, among other fluctuating properties. These fluctuations are dominated by the equilibrium distribution and depend weakly on other process properties.

\end{abstract}
 
\maketitle

\section{Introduction}

Repeated interaction schemes, a.k.a collisional models~\cite{rep-int,rep-int2,Attal2,Karevski,rep-int3,rep-int4}, have played a vital role in the development of quantum optics~\cite{maser1,maser2,maser3,maser4} and the rapid evolution of quantum thermodynamics~\cite{Qth1,Qth2,Qth3,Qth4,Qth5}. The idealized and straightforward formalism has been crucial to designing and understanding quantum devices such as information engines~\cite{info1,info2,info3,Esposito rep. int.}, heat engines~\cite{Qth2, rep-int-engine,DenzlerLutz,enginePhil,PreB}, and quantum batteries~\cite{QBRepInt,QBRepInt2,BarraBattery,firstQB,quantacell1,quantacell2,campisibatt,qb1,qb2,qb3}.
Recently, it was realized that the framework can be extended to deal with macroscopic reservoirs~\cite{PreB2,PreB}, expanding the reach of applications in quantum thermodynamics. 
For comprehensive reviews of the method and its applications, see~\cite{rev1} and~\cite{rev2}. 

In the simplest scenario, many copies of an auxiliary system in the Gibbs equilibrium thermal state interact sequentially with a system of interest. Each interaction step is described by a completely-positive trace-preserving (CPTP) map~\cite{BreuerBook}. The repeated interaction process corresponds to concatenations of the map, which eventually will bring the system to a nonequilibrium steady-state or an equilibrium state. In equilibrium, heat does not flow to the environment, and entropy is not produced. They are not sustained by work. When the repeated interaction brings the system to an equilibrium state, we say that we iterate a map with equilibrium.

The thermodynamic quantities characterizing the process can be considered the average over stochastic versions of them defined on trajectories. These quantities satisfy fluctuation theorems~\cite{MHPPRE2015,HPNJP2013,Campisi,MHPPRX2018} that reveal essential properties of the process. Particular attention has been drawn to efficiency fluctuations in different classical~\cite{ef1,ef2,ef3,ef4,ef5,ef6,ef7,ef8,ef9,ef10,ef11} or quantum~\cite{DenzlerLutz} engines due to their remarkable properties. 

This paper will study a quantum battery charged by a repeated interaction process. A quantum battery is a system that stores energy. The battery's charge is characterized by its ergotropy~\cite{ergotropy}, i.e., the maximum amount of energy extracted with a unitary process. Once the energy is removed, the battery is recharged with a repeated interaction process that starts from the discharged state. In this way, we have a working cycle, and we analyze its thermodynamics. The most straightforward charging protocol considers the auxiliary systems in a nonequilibrium state. However, the process of sustaining the charged state is dissipative. Ref~\cite{BarraBattery} proposed a different kind of quantum battery. The charged state corresponds to the equilibrium state of the process. Therefore the charge is preserved without dissipation and provided by the work done in the recharging stage.

We will study the efficiency fluctuations of the charging process.
We will see that the equilibrium state of the battery determines its statistics. We will illustrate this in two examples. We will also study work fluctuations in the recharging stage. We discuss equilibrium fluctuations and the difference between fluctuations in the equilibrium states of battery systems and the Gibbs equilibrium state. To characterize the fluctuations of the quantum system, we consider the two-point measurement scheme~\cite{esposito-mukamel-RMP}. Evaluating the fluctuations requires detailed information about the bath and the process. However, a simplification arises because we deal with maps with equilibrium. 

The remainder of this article is organized as follows. In section \ref{traj.map.sec}, we review the thermodynamics for CPTP maps emphasizing the results for maps with equilibrium. Then, in section \ref{sec:battery}, we introduce our system of study, namely the equilibrium quantum battery proposed in~\cite{BarraBattery}. Section \ref{sec.fluct} discusses the stochastic versions of the thermodynamic equalities and laws, emphasizing the results for maps with equilibrium again. Subsequently, in section \ref{sec:ejemplos}, we evaluate these fluctuations in two illustrative examples.
We conclude this article in section \ref{secCONC}.

\section{Thermodynamic description for completely positive trace-preserving maps}
\label{traj.map.sec}

Consider a system $S$ and a system $B$ that jointly evolve under the unitary  $U=e^{-i\frac{\tau}{\hbar}(H_S+H_B+V)}$.
The Hamiltonians $H_S$ and $H_B$ of $S$ and $B$ respectively are constant in time. The coupling between $S$ and $B$ during the time interval $(0,\tau)$ is given by the interaction energy $V$ and vanishes for $t<0$ and $t>\tau$. 

Initially, $S$ and $B$ are uncorrelated, i.e., their density matrix is the tensor product of the respective density matrices  $\rho_{\rm tot}=\rho_S\otimes \omega_\beta(H_B)$, where $\omega_\beta(H_B)=\frac{e^{-\beta H_B}}{Z_B}$ is the Gibbs thermal state for $B$ with $\beta$ the inverse temperature, and $Z_B={\rm Tr} \,e^{-\beta H_B}.$
After the lapse of time $\tau$ the initial state $\rho_{\rm tot}$ changes to a new state,
\begin{equation}
\label{unitary}
\rho'_{\rm tot}=U \left(\rho_S\otimes\omega_\beta(H_B)\right) U^\dag.
\end{equation}
In the following, we denote $\rho_S'={\rm Tr}_B\rho_{\rm tot}'$ and $\rho_B'={\rm Tr}_S\rho_{\rm tot}'$, where ${\rm Tr}_X$ is the partial trace over subsystem $X$. By tracing out $B$, one obtains a CPTP map $\mathcal{E}$ for the system $S$ evolution
\begin{equation}
\rho'_S=\mathcal{E}(\rho_S)={\rm Tr}_B\left[U \left(\rho_S\otimes\omega_\beta(H_B)\right) U^\dag\right].
\label{CPTP}
\end{equation}

The energy change of $S$
\begin{equation}
\Delta E={\rm Tr}[H_S(\rho'_S-\rho_S)],
\label{Av.Energy}
\end{equation}
can be written as the sum of
\begin{equation}
Q={\rm Tr}[H_B(\omega_\beta(H_B)-\rho'_B)],
\label{Av.Heat}
\end{equation}
and
\begin{equation}
W={\rm Tr}[(H_S+H_B)(\rho'_{\rm tot}-\rho_{\rm tot})],
\label{Av.Work}
\end{equation} 
satisfying the first law $\Delta E=W+Q$. Note that $Q$ is minus the energy change of $B$, we call it heat and $W$ is the energy change of the full $S+B$ system, we call it the switching work because it is due to the energy cost of turning on and off the interaction $V$ at the beginning and end of the process respectively~\cite{Barra2015,Chiara}.

Consider the von Neumann entropy change 
\begin{equation}
\Delta S_{\rm vN}=-{\rm Tr}[\rho_S'\ln\rho_S']+{\rm Tr}[\rho_S\ln\rho_S]
\label{Av.Ent}
\end{equation}
of system $S$ and the heat $Q$ given in Eq.~(\ref{Av.Heat}). The entropy production 
$\Sigma=\Delta S_{\rm vN}-\beta Q$,
is also given by~\cite{esposito-NJP}
\begin{equation}
\Sigma=D(\rho_{\rm tot}'||\rho_S'\otimes \omega_\beta(H_B))\geq 0,
\label{Av.Ent.Prod}
\end{equation}
with 
 $D(a||b)\equiv\Tr[a\ln a] - \Tr[a \ln b].$ The inequality in Eq.\eqref{Av.Ent.Prod} corresponds to the second law.
Note that system $B$ does not need to be macroscopic, nevertheless, we will call it the bath. 

As in standard thermodynamics, analyzing the process $\rho_S\to\rho_S'={\mathcal E}(\rho_S)$, in terms of $\Delta E=W+Q$ and $\Sigma=\Delta S_{\rm vN}-\beta Q\geq 0$
with the quantities given in Eqs.~\eqref{Av.Energy},~\eqref{Av.Heat},~\eqref{Av.Work},~\eqref{Av.Ent}, and~\eqref{Av.Ent.Prod} is very useful.
Note that for their evaluation, particularly for the work, Eq.~(\ref{Av.Work}), and entropy production, Eq.~(\ref{Av.Ent.Prod}), we need to know the full state $\rho_{\rm tot}'$.

\subsection{ Maps with thermodynamic equilibrium}

In a repeated interaction process, one concatenates $L$ CPTP maps $\mathcal{E}^L\equiv\mathcal{E}\circ\cdots\circ\mathcal{E}(\cdot)$ to describe a sequence of  evolutions of a system coupled to a heat bath for a given lapse of time $\tau$. With each map $\mathcal{E}$,  a new fresh bath is introduced that exchanges heat with the system during the time that the interaction is turned on. The concatenated map $\mathcal{E}^L$  is also a CPTP map. The total work performed is the sum of the work done switching on and off the interaction energy with each bath. Similarly, the total heat is the sum of the heat exchanged with each bath. 

Let us assume that the map ${\mathcal E}$ has an attractive invariant state $\bar{\rho}$ defined as \[
\lim_{L\to\infty}{\mathcal E}^L(\rho_S)=\bar{\rho}, \,\forall \rho_S,
\]
and $\bar{\rho}={\mathcal E}(\bar{\rho})$.
The process $\bar{\rho}\to\mathcal{E}(\bar{\rho})$ is thermodynamically characterized by $\Delta S_{\rm vN}=0=\Delta E$, see Eq.~(\ref{Av.Energy}) and Eq.~(\ref{Av.Ent}). If the entropy produced by the action of the map ${\mathcal E}$ on $\bar{\rho}$ is $\Sigma>0$, then we say that the invariant state is a non-equilibrium steady state. The invariant state is an {\it equilibrium state} if $\Sigma=0$, i.e., if the entropy production, Eq.~(\ref{Av.Ent.Prod}), vanishes by the action of $\mathcal E$ on $\bar{\rho}$. Maps with these particular states are called maps with equilibrium~\cite{StochPRE,StochPRE2}.  

According to Eq.~(\ref{Av.Ent.Prod}), $\Sigma=0$ for the steady state $\bar{\rho}$ if and only if $\bar{\rho}\otimes\omega_\beta(H_B)=U \left(\bar{\rho}\otimes\omega_\beta(H_B)\right) U^\dag$.  
Equivalently, 
if the unitary $U$  in Eq.~(\ref{unitary}) satisfies $[U,H_0+H_B]=0$, where $H_0$ is an operator in the Hilbert space of the system, then the product state $\omega_\beta(H_0)\otimes\omega_\beta(H_B)$, with $\omega_\beta(H_0)=\frac{e^{-\beta H_0}}{Z_0}$, where $Z_0={\rm Tr}[e^{-\beta H_0}]$, is invariant under the unitary evolution in Eq.~(\ref{unitary}) and $\bar{\rho}=\omega_\beta(H_0)$ is an equilibrium state for the map in Eq.~(\ref{CPTP}).

It follows from $[U,H_0+H_B]=0$ that 
the heat, Eq.\eqref{Av.Heat} and work, Eq.\eqref{Av.Work} simplify to
\begin{equation}
Q={\rm Tr}[H_0(\rho_S'-\rho_S)]
\label{Eq.prop}
\end{equation}
and
\begin{equation}
W={\rm Tr}_S[(H_S-H_0)(\rho_S'-\rho_S)].
\label{Av.Work.Eq}
\end{equation}
The entropy production also reduces to an expression that does not involve the state of the bath.
Indeed, we obtain 
\begin{equation}
\Sigma
=D(\rho_S||\omega_\beta(H_0))-D(\rho_S'||\omega_\beta(H_0)),
\label{epthermal}
\end{equation}
which is positive due to the contracting character of the map~\cite{BreuerBook}. 
 The averaged thermodynamic quantities for a map with equilibrium are only determined by the properties of the system of interest.

If $H_0=H_S$, then the map is called thermal~\cite{terry2,terry3}.
The equilibrium state is the Gibbs state $\omega_\beta(H_S)=e^{-\beta H_S}/Z_S$ with $Z_S={\rm Tr}[e^{-\beta H_S}]$, and the agent is passive because $W=0$ for every initial state $\rho_S$, see Eq.~(\ref{Av.Work.Eq}).

When $H_0\neq H_S$, an active external agent has to provide (or extract) work to perform the map on a state $\rho_S$.  However, once the system reaches the equilibrium state $\omega_\beta(H_0)$, the process $\omega_\beta(H_0)\to{\mathcal E}(\omega_\beta(H_0))=\omega_\beta(H_0)$ is performed with $W=0$, see Eq.~(\ref{Av.Work.Eq}), and $\Sigma=0$.

Let us end this section with the following remark. Since the total evolution operator $U=e^{-i\frac{\tau}{\hbar}(H_S+H_B+V)}$ is time-independent, the equilibrium condition is satisfied by finding $H_0$ and $V$ such that $[H_0,H_S]=0$ and $[H_0+H_B,V]=0$~\cite{BarraBattery}. 
In this case, $H_S$ and $H_0$ share the same eigenbasis. To simplify the discussion of fluctuations, we consider non-degenerate eigenenergies. We denote the eigensystems as
\[
H_S\ket{n}=E_n\ket{n},\quad H_0\ket{n}=E^0_n\ket{n}.
\]
whith increasing order $E_1< E_2<\cdots< E_N$ for the eigenenergies. The eigenvalues $E_n^0$ are not necessarily ordered but there is always a permutation that we call $\pi$ of $(1,\ldots,N)\to (\pi_1,\ldots,\pi_N)$ such that $E_{\pi_1}^0\leq \cdots\leq E_{\pi_N}^0.$

\section{The battery}
\label{sec:battery}
As is well known, the Gibbs state $\omega_\beta(H_S)$ is passive, i.e., one can not decrease (extract) its energy with a unitary operation~\cite{passivity1,passivity2}. This is not true for the equilibrium state 
\begin{equation}
\label{cargado}
\omega_\beta(H_0)=\sum_n \frac{e^{-\beta E_n^0}}{Z_0}\ket{n}\bra{n},
\end{equation} 
if a pair $(j,k)$ exists such that $(E_j-E_k)(E_j^0-E_k^0)<0$. In that case, the unitary operator $u$ with matrix elements $u_{ij}=\braket{i|u|j}=\delta_{\pi_i,j}$ extracts the ergotropy~\cite{ergotropy}
\begin{equation}
\label{ergotropy1}
{\mathcal W}[\omega_\beta(H_0)]=\sum_{n=1}^N (E_{\pi_n}-E_n)\frac{e^{-\beta E_{\pi_n}^0}}{Z_0}> 0,
\end{equation}
where $\pi$ is the permutation that orders $E_n^0$ increasingly.  

Once the ergotropy is extracted, the system is left in the passive state 
\begin{equation}
\label{pasivo}
\sigma_{\omega_\beta(H_0)}=u \omega_\beta(H_0) u^\dag=\sum_{n=1}^N\frac{e^{-\beta E_{\pi_n}^0}}{Z_0}\ket{n}\bra{n}.
\end{equation}
An equilibrium quantum battery was proposed in~\cite{BarraBattery} based on that observation.
The system is driven by a repeated interaction process described by a map ${\mathcal E}$ with equilibrium $\omega_\beta(H_0)$. Once the equilibrium is reached, it is kept with no cost ($ W=0$), energy does not leak from it, and the battery's charge, characterized by the ergotropy ${\mathcal W}[\omega_\beta(H_0)]$ is preserved. Equilibrium states with ergotropy are called active.

The thermodynamic cycle is as follows: The battery starts in the active equilibrium state, then the ergotropy \eqref{ergotropy1} is extracted, leaving the battery in the passive state \eqref{pasivo}
from which the repeated interaction process $\lim_{L\to\infty}{\mathcal E}^L(\sigma_{\omega_\beta(H_0)})$ recharges it.  
As a consequence of the second law, the recharging work $W_R={\rm Tr}_S[(H_S-H_0)(\omega_\beta(H_0)-\sigma_{\omega_\beta(H_0)})]$ is never smaller that the extracted ergotropy. In this way, the thermodynamic efficiency $0\leq \eta_{\rm th}\equiv  {\mathcal W}[\omega_\beta(H_0)]/W_R\leq 1$ characterizes the operation of the device.

\section{Fluctuations}
\label{sec.fluct}

\subsection{repeated interaction for a map with equilibrium}

The thermodynamic quantities in Eqs.\eqref{Av.Energy}, \eqref{Av.Heat}, \eqref{Av.Work}, \eqref{Av.Ent} and \eqref{Av.Ent.Prod} were obtained as the average over their stochastic versions defined over trajectories using a two-point measurement scheme in~\cite{StochPRE}. Since all interesting density matrices $\omega_\beta(H_S), \omega_\beta(H_0)$ and $\sigma_{\omega_\beta(H_0)}$ are diagonal in the system energy basis, we need only projective energy measurement in this work.

A trajectory $\gamma=\{n;i_1,j_1,\ldots,i_L,j_L;m\}$ for the recharging process is defined by the initial and final, $\epsilon_{i_k}$ and $\epsilon_{k_k}$, energy result for each auxiliary thermal system and  $E_n$ and $E_m$ for the system. According to the two-point measurement scheme~\cite{esposito-mukamel-RMP}, its probability is
\begin{equation}
P^{(L)}_\gamma=|\langle j_1\cdots j_L m|U_L\cdots U_1|i_1\cdots i_L n\rangle|^2 \frac{e^{-\beta \sum_{k=1}^L\varepsilon_{i_k}}}{Z_B^L}p_{\rm ini}(n),
\label{psm1}
\end{equation}
where $p_{\rm ini}(n)$ is the probability that the initial state of the system is $\ket{n}$, see Appendix~\ref{sec:appendix}.
We now associate the stochastic thermodynamic quantities to these trajectories.
The stochastic heat flow to the system $q_\gamma$ 
corresponds to the negative energy change of the bath, i.e., $q_\gamma=\sum_{k=1}^L(\varepsilon_{i_k}-\varepsilon_{j_k})$.
According to the first law of stochastic thermodynamics~\cite{HPNJP2013}, the stochastic work is given by
\begin{equation}
w_\gamma=\Delta e_\gamma-q_\gamma,
\label{w-gamma}
\end{equation}
where $\Delta e_\gamma=E_m-E_n$ is the stochastic energy change. These fluctuating quantities are studied through their distributions
\begin{equation}
 p^{(L)}_w(x)=\sum_\gamma \delta(x-w_\gamma)P^{(L)}_\gamma,\quad
p^{(L)}_{\Delta e}(x)=\sum_\gamma \delta(x-\Delta e_\gamma)P^{(L)}_\gamma,\quad
p^{(L)}_q(x)=\sum_\gamma \delta(x-q_\gamma)P^{(L)}_\gamma,
\label{p(w)}
\end{equation}
and, as for the averaged thermodynamic quantities, we need information on the state of the whole system to evaluate them. However, for maps with equilibrium, a stochastic trajectory is determined by the pair $\gamma=\{n,m\}$, see Appendix~\ref{sec:appendix}. Consequently these formulas simplify and become, $q_\gamma=E^0_m-E^0_n, w_\gamma=E_m-E_m^0-(E_n-E_n^0)$ with the distributions 
\begin{equation} \label{ecc:dist energ eq. map}
p^{(L)}_{\Delta e}(x)=\sum_{n,m} \delta(x-[E_m-E_n])P^{(L)}_{n\to m},
\end{equation}
\begin{equation} \label{ecc:dist trabajo eq. map}
p^{(L)}_w(x)=\sum_{n,m} \delta(x-[(E_m-E_m^0)-(E_n-E_n^0)])P^{(L)}_{n\to m},
\end{equation}
\begin{equation} \label{ecc:dist calor eq. map}
p^{(L)}_q(x)=\sum_{n,m} \delta(x-[E_m^0-E_n^0])P^{(L)}_{n\to m},
\end{equation}
and the trajectory probability
\begin{equation}
\label{transprob}
P^{(L)}_{n\to m}=\bra{m}{\mathcal E}^L(\ket{n}\bra{n})\ket{m}p_{\rm ini}(n)=(T^L)_{m|n}\,p_{\rm ini}(n),
\end{equation} 
in terms of the initial probability $p_{\rm ini}(n)$ and of the $L$ power of the stochastic matrix $T_{m|n}=\bra{m}{\mathcal E}(\ket{n}\bra{n})\ket{m}$.

The averages $\int x p^{(L)}_{\Delta e}(x)dx,\int xp^{(L)}_w(x)dx,\int xp^{(L)}_q(x)dx$ reproduce Eqs.\eqref{Av.Energy}, \eqref{Eq.prop},  \eqref{Av.Work.Eq} with $\rho'_S={\mathcal E}^L(\rho_S)$ and $\rho_S=\sum_n p_{\rm ini}(n)\ket{n}\bra{n}$. 

\subsection{Fluctuations in the equilibrium state}
\label{sec.eq.fluct}

As noted before, all averaged thermodynamic quantities $\Delta E=\Delta S=\Sigma=W=Q=0$ vanish for a process in equilibrium. So, on average, the process $\omega_\beta(H_0)\to{\mathcal E}(\omega_\beta(H_0))=\omega_\beta(H_0)$ has no energy cost. However, if $H_0\neq H_S$, the agent is still active due to non-vanishing work fluctuations. For thermal maps $H_0=H_S$, and Eq.\eqref{ecc:dist trabajo eq. map} gives $p^{(L)}_w(x)=\delta(x)$. The external agent is truly passive.  

To analyze equilibrium fluctuations, we use Eqs.\eqref{ecc:dist energ eq. map}, \eqref{ecc:dist trabajo eq. map} and \eqref{ecc:dist calor eq. map} with $p_{\rm ini}(n)=\frac{e^{-\beta E_{n}^0}}{Z_0}$. 

\subsection{recharging process}
\label{sec.rech.fluct}

Since the recharging process starts from $\sigma_{\omega_\beta(H_0)}$, we take $p_{\rm ini}(n)=e^{-\beta E_{\pi_n}^0}/Z_0$, see Eq.\eqref{pasivo}, in the distributions Eqs.\eqref{ecc:dist energ eq. map}, \eqref{ecc:dist trabajo eq. map}, and \eqref{ecc:dist calor eq. map}.

Since the charged state $\omega_\beta(H_0)$ is reached asymptotically, we take $L\to\infty$ to charge the battery fully. When $L$ is finite, we speak of partial recharging. However, in that case, we do not have a cyclic engine because the passive state associated with ${\mathcal E}^L(\sigma_{\omega_\beta(H_0)})$ is not $\sigma_{\omega_\beta(H_0)}$. In the limit $L\to \infty$ we have a well defined cycle. 

Moreover, since ${\mathcal E}$ has a unique equilibrium state, we will find that $T$ is a regular stochastic matrix~\cite{Feller} implying that $\lim_{L\to\infty}(T^L)_{m|n}=e^{-\beta E^0_m}/Z_0,\forall n$.  
Therefore, the limit in Eq.\eqref{transprob}
\begin{equation}
\label{Pcharge}
P^{(\infty)}_{n\to m}=p_{\rm ini}(n)e^{-\beta E^0_m}/Z_0=e^{-\beta (E_{\pi_n}^0+E^0_m)}/Z_0^2,
\end{equation} 
is independent of the map's details. Interestingly, the rate of convergence of $T^L$ to the equilibrium distribution depends on the map $\mathcal E$ parameters. We discuss later the fluctuations of a concatenated process $\mathcal E^L$ with finite $L$.

The average of the stochastic energy change in the recharging process
\begin{equation}
\label{ergotropy2}
\langle\Delta e_\gamma\rangle^{(\infty)}\equiv \sum_{n,m} (E_m-E_n)P^{(\infty)}_{n\to m}=\Tr[H_S(\omega_\beta(H_0)-\sigma_{\omega_\beta(H_0)})]={\mathcal W}(\omega_\beta(H_0))
\end{equation}
 is the ergotropy. The average stochastic work 
 \begin{equation}
 \label{WR2}
 \langle w_\gamma\rangle^{(\infty)}\equiv \sum_{n,m} ((E_m-E_m^0)-(E_n-E_n^0))P^{(\infty)}_{n\to m}=\Tr[(H_S-H_0)(\omega_\beta(H_0)-\sigma_{\omega_\beta(H_0)})]=W_R
 \end{equation}
 is the recharging work.

\subsection{extracting process}
The extracting process is also fluctuating when we measure the battery's energy in the charged state and the discharged state. We call $\kappa$ the stochastic trajectory in the ergotropy extracting process and $\varpi_\kappa$ the stochastic extracted energy. The probability $p_\kappa$ of $\kappa=(m',n)$ is the product of  the transition probability from $\ket{m'}$ to $\ket{n}$ under the permutation $u$, $P^{\rm ext}_{m'\to n}=|\braket{n|u|m'}|^2=\delta_{\pi_n,m'}$, with the initial probability $e^{-\beta E_{m'}^0}/Z_0$, see Eq.\eqref{cargado}.  
The averaged extracted energy,
 \begin{equation}
 \langle{\varpi}_{\kappa}\rangle=\sum_\kappa \varpi_\kappa p_\kappa=\sum_{m',n}(E_{m'}-E_n)P^{\rm ext}_{m'\to n}\frac{e^{-\beta E_{m'}^0}}{Z_0}=\sum_{n}(E_{\pi_n}-E_n)\frac{e^{-\beta E_{\pi_n}^0}}{Z_0}={\mathcal W}(\omega_\beta(H_0))
 \label{erg-av}
 \end{equation}
is the ergotropy.

Eq.\eqref{ergotropy2} and Eq.\eqref{erg-av} show the cycle's consistency, where two processes, recharging ($\gamma$) and extracting ($\kappa$) connect the same states, $\omega_\beta(H_0)$ and 
$\sigma_{\omega_\beta(H_0)}$.

\subsection{ Fluctuating efficiency for the cycle}
\label{sec.eff.fluct}

In terms of Eq.\eqref{erg-av} and Eq.\eqref{WR2} we have the thermodynamic efficiency $\eta_{\rm th}=\frac{{\mathcal W}}{W_R}=\frac{\langle\varpi_\kappa\rangle}{\langle w_\gamma\rangle^{(\infty)}}$. 
 
As the thermodynamic efficiency is the ratio of the ergotropy over the recharging work, the fluctuating efficiency~\cite{DenzlerLutz} should be the ratio of their fluctuating equivalents. The fluctuating extracted energy is $\varpi_\kappa=E_{m'}-E_n$, and the fluctuating work is $w_\gamma=E_m-E_m^0-(E_n-E_n^0)$. Therefore, we define the fluctuating efficiency as  
\begin{equation}
\label{flucteff}
\eta_{\gamma\kappa}=\frac{\varpi_\kappa}{w_\gamma}=\frac{E_{m'}-E_n}{E_{m}-E_{m}^0-(E_n-E_n^0)}.
\end{equation}
Given the extracting trajectory $\kappa$, the probability of the recharging trajectory $\gamma$ is $P^{\rm ext}_{m'\to n}P^\infty_{n\to m}$. Thus, the joint probability 
for the processes $\kappa$ and $\gamma$ is
\[
p_{\gamma \kappa}=\frac{e^{-\beta E_{m'}^0}}{Z_0}P^{\rm ext}_{m'\to n}P^\infty_{n\to m}=\frac{e^{-\beta E_{m'}^0}}{Z_0}\delta_{\pi_n,m'}\frac{e^{-\beta E^0_m}}{Z_0},
\]
and the distribution of the fluctuating efficiency is
\begin{equation} \label{ecc:dist efficiency eq. map prima}
p_\eta(x)=\sum_{\gamma,\kappa}\delta(x-\eta_{\gamma\kappa})p_{\gamma\kappa}=
\sum_{n,m} \delta\left(x-\frac{E_{\pi_n}-E_n}{E_{m}-E_{m}^0-(E_n-E_n^0)}\right)\frac{e^{-\beta (E^0_m+E_{\pi_n}^0)}}{Z_0^2}.
\end{equation}
To simplify the notation we write this as
\begin{equation}
\label{etadistrinm}
p_\eta(x)
=\sum_{n,m} \delta\left(x-\eta_{nm}\right)P_{n\to m},
\end{equation}
with 
\begin{equation}
\label{flucteff2}
\eta_{nm}=\frac{E_{\pi_n}-E_n}{E_{m}-E_{m}^0-(E_n-E_n^0)},\quad \text{and}\quad P_{n\to m}=\frac{e^{-\beta (E^0_m+E_{\pi_n}^0)}}{Z_0^2}.
\end{equation}
The probability $P_{n\to m}$ corresponds Eq.\eqref{Pcharge} and we omit the superscript.

Trajectories with $w_\gamma=0$ have $\eta_{\gamma\kappa}=\infty$. Therefore the average $\langle\eta_{\gamma\kappa}\rangle$ does not always exist, and if it does, $\eta_{\rm th}\neq \langle\eta_{\gamma\kappa}\rangle$, 
unless the stochastic work and efficiency are uncorrelated.
In fact, 
$\langle \eta_{\gamma\kappa} w_\gamma\rangle=\langle \varpi_\kappa\rangle={\mathcal W}$. So only if 
$\langle \eta_{\gamma\kappa} w_\gamma\rangle
=\langle \eta_{\gamma\kappa}\rangle W_R$ 
we have $\langle\eta_{\gamma\kappa}\rangle=\eta_{\rm th}$. The thermodynamic and fluctuating efficiency can be very different.

The following section discusses efficiency fluctuations for the cycle, heat and work fluctuations for the recharging process and equilibrium fluctuations in two examples.

\section{Examples}
\label{sec:ejemplos}

We illustrate our results in two simple examples. 
The first example is a single-qubit battery that we use to discuss equilibrium fluctuations (section~\ref{sec.eq.fluct}). The second example is a  
two-qubit battery where we compute heath and work distributions in a partial recharging process (section~\ref{sec.rech.fluct}). 
In both, we compute the fluctuating efficiency distribution (section~\ref{sec.eff.fluct}).

\subsection{Single-qubit battery}

An interesting protocol, with $H_0=-H_S$, was discussed in~\cite{BarraBattery} for a system $S$ interacting with systems $B$, which are copies of $S$. 
The corresponding process ${\mathcal E}$ has the remarkable equilibrium state 
\[
\omega_\beta(-H_S)=\sum_{n=1}^N \frac{e^{\beta E_n}}{Z_+}\ket{n}\bra{n},
\]
with $Z_+=\Tr[e^{+\beta H_S}]$ between a system in the state $\omega_\beta(-H_S)$ with copies of itself in the Gibbs state $\omega_\beta(H_S)$.

In this subsection, we consider the battery $S$ and auxiliary systems $B$ identical qubits; i.e., the battery Hamiltonian is $H_S=(h/2) \sigma_S^z$, and the baths Hamiltonians are $H_B=(h/2) \sigma_B^z$, with $h>0$. Hereafter, $\sigma^x,\sigma^y$, and $\sigma^z$ are Pauli matrices.

The coupling between the system and the bath qubit is 
\[
V=a(\sigma_S^+\sigma_B^++\sigma_S^-\sigma_B^-),
\]
with $\sigma^\pm=(\sigma^x\pm \sigma^y)/2$, and is such that $[\sigma_B^z-\sigma_S^z, V]=0$, i.e., $H_0=-H_S$. 

In the basis defined by $\sigma^z\ket{\uparrow}=\ket{\uparrow}$ and $\sigma^z\ket{\downarrow}=-\ket{\downarrow}$, the eigenvalues and eigenvectors of $H_S$ and $H_0$ are  
\begin{align}
E_2&=h/2,        & E^0_2&=-h/2, &       \ket{2}&=\ket{\uparrow}\\
E_1&=-h/2,       & E^0_1&=h/2, &     \ket{1}&=\ket{\downarrow}
\end{align}
and the ordering permutation is $(\pi_1,\pi_2)=(2,1)$. 
Thus, on the above basis, the equilibrium state is
\[
\omega_{\beta}(H_0)=\omega_{\beta}(-H_S)
=\frac{e^{\beta \frac{h}{2}} }{Z}\ket{2}\bra{2}+\frac{e^{-\beta \frac{h}{2}}}{Z}\ket{1}\bra{1},
\]
and the passive state for the system is 
\[
\sigma_{\omega_\beta(H_0)}=\omega_{\beta}(H_S)=\frac{e^{-\beta \frac{h}{2}} }{Z}\ket{2}\bra{2}+\frac{e^{\beta \frac{h}{2}}}{Z}\ket{1}\bra{1},
\]
where $Z=Z_+=2\cosh(\beta h/2).$
The ergotropy of the battery in the equilibrium state $\omega_\beta(-H_S)$ is ${\mathcal W}=h\tanh \beta h/2$. From Eqs.\eqref{ergotropy2} and  \eqref{WR2}, we see that the thermodynamic efficiency of the process is $\eta_{\rm th}=1/2,$ independent of the inverse temperature $\beta$.

The recharging process in this single-qubit battery (1Q) is determined by the stochastic matrix (see Eq.\eqref{transprob})
\begin{equation}
\label{Tqubitbattery}
T_{1Q}=\left(\begin{array}{cc}
1-\frac{e^{\beta \frac{h}{2}}}{Z}g(a,h) & \frac{e^{-\beta\frac{h}{2}}}{Z}g(a,h)\\
\frac{e^{\beta\frac{h}{2}}}{Z}g(a,h) & 1-\frac{e^{-\beta \frac{h}{2}}}{Z}g(a,h)
\end{array}\right)
\end{equation}
where $g(a,h)=\frac{a^2 \sin^2(\tau\sqrt{h^2+a^2}/\hbar)}{h^2+a^2}$ and $Z=e^{\beta \frac{h}{2}}+e^{-\beta \frac{h}{2}}$. It is a regular stochastic matrix if $g(a,h)\neq 0$.

\subsubsection{fluctuating efficiency}
The fluctuating efficiency (see Eq.\eqref{flucteff2}) takes the values

\[
\eta_{11}=\eta_{22}=\infty,\quad
\eta_{12}=\eta_{21}=\frac{1}{2}
\]
Its distribution Eq.\eqref{etadistrinm} is
\[
p_\eta(x)=\delta(x-\infty)\mathrm{P}_\infty+\delta\left(x-\frac{1}{2}\right)\mathrm{P}_\frac{1}{2}
\]
with
\begin{equation}
\label{peff1q}
\mathrm{P}_\infty=P_{1\to 1}+P_{2\to 2}=\frac{2}{Z^2},\quad
\mathrm{P}_\frac{1}{2}=P_{1\to2}+P_{2\to 1}=\frac{e^{\beta h}+e^{-\beta h}}{Z^2}
\end{equation}
The explicit formulas at the right follow from Eq.\eqref{flucteff2}
which is valid if $g(a,h)\neq 0$ in $T_{1Q}$.

\begin{figure}[H] 
\centering
\includegraphics[width=0.4\textwidth]{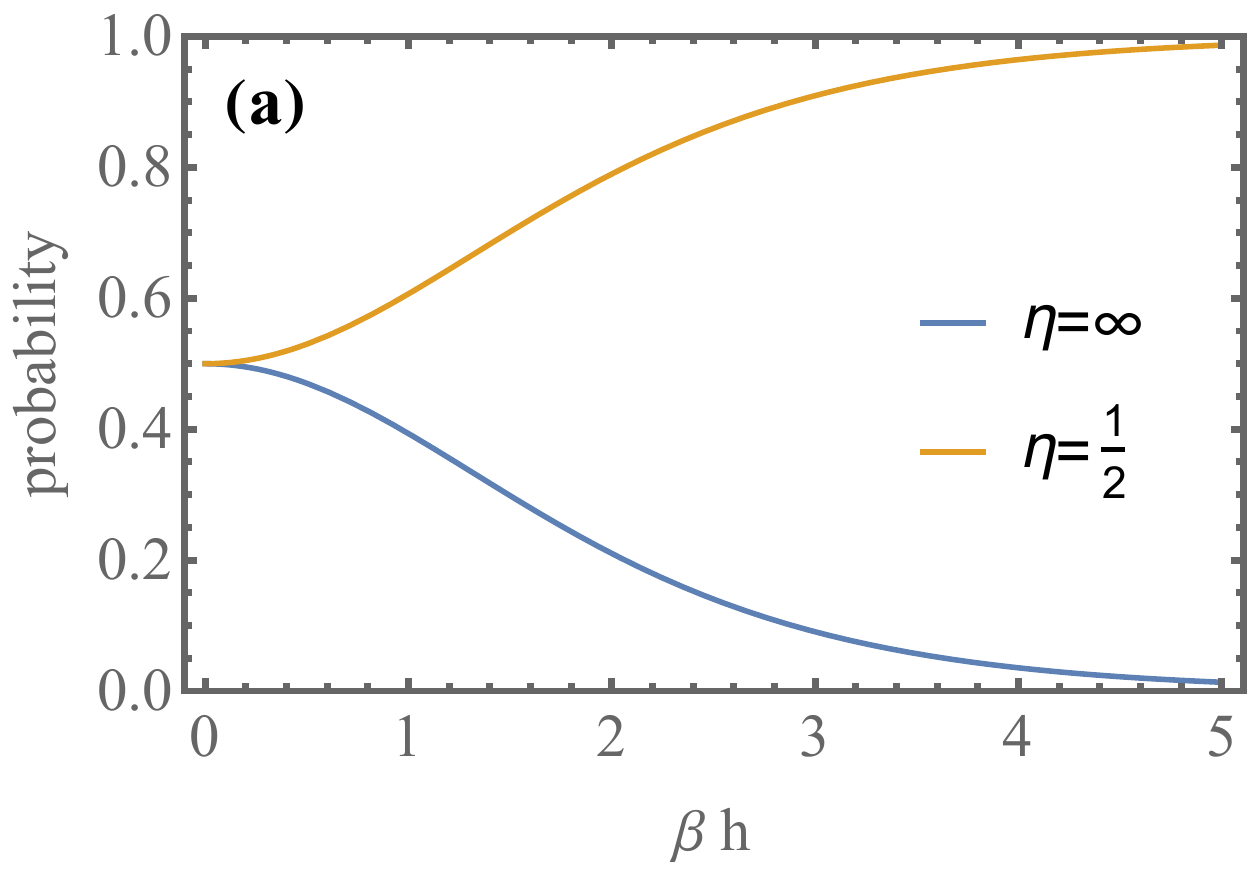}
\includegraphics[width=0.4\textwidth]{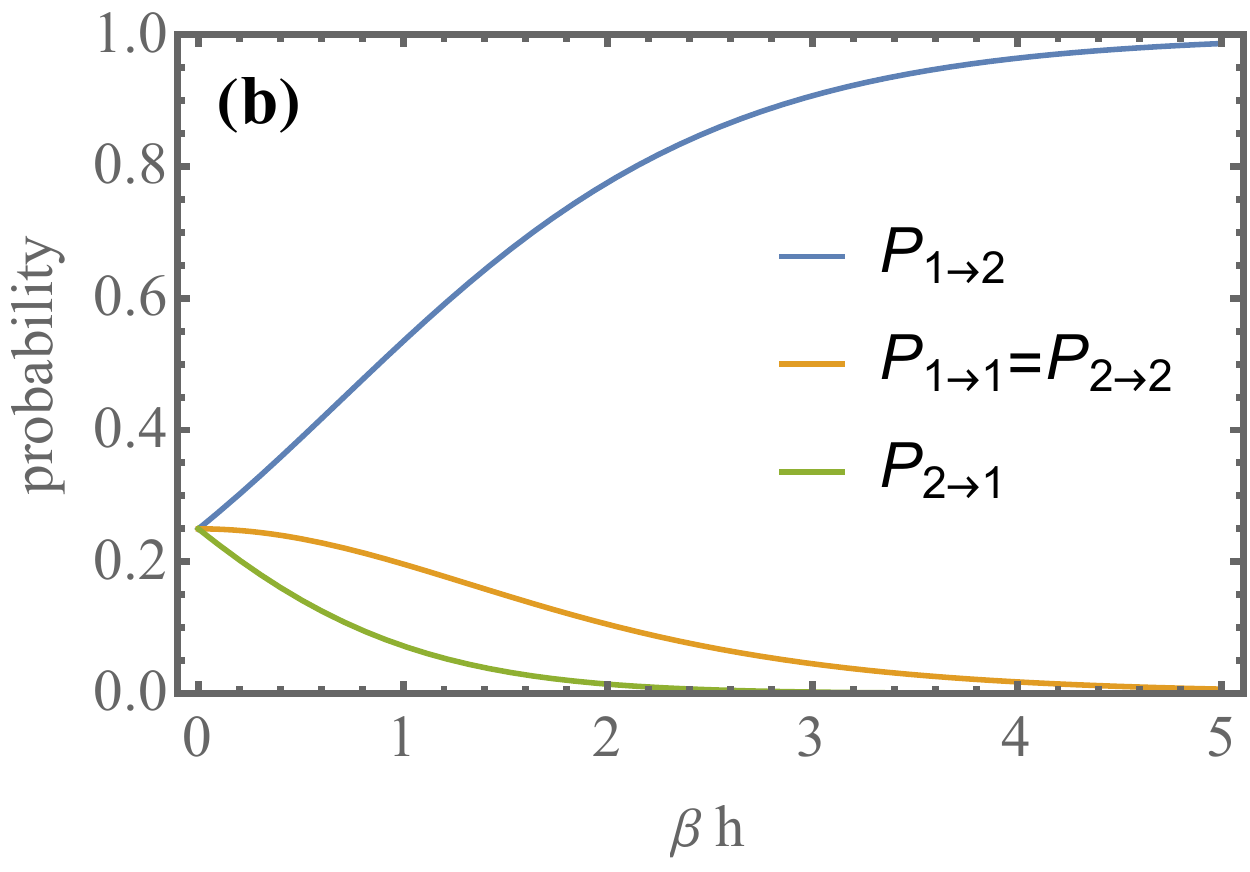}
    \caption{For the 1-qubit battery (a) Plots of $\mathrm{P}_\eta$ (Eq.\eqref{peff1q}) as function of $\beta h$. (b) Plots of $P_{n\to m}$ given by Eq.\eqref{flucteff2} for the single-qubit battery}
    \label{figura qubit}
\end{figure}

In Figure~\ref{figura qubit}a, we depict the probabilities $\mathrm{P}_\eta$ as functions of $\beta h$ and see that for $\beta h\gg1$ with probability 1, the fluctuating efficiency equals the thermodynamic efficiency $1/2$, because, as we see in Figure~\ref{figura qubit}b, 
 $P_{1\to 2}\to 1$ reflecting the charging character of the process. 

\subsubsection{equilibrium fluctuation}

Let us analyze the fluctuations when maintaining the charged state, i.e., those of the process $\omega_\beta(H_0)\to{\mathcal E}^L(\omega_\beta(H_0))=\omega_\beta(H_0)$, see section~\ref{sec.eq.fluct}.  
As we can verify in the examples above, and as shown in~\cite{StochPRE}, the transition matrices $T$ for maps with equilibrium satisfy the detailed balance condition $T_{m|n}e^{-\beta E_n^0}=T_{n|m}e^{-\beta E_m^0}$. From this, it is simple to show that $P^{(L)}_{n\to m}=P^{(L)}_{m\to n}$ with $p_{\rm ini}(n)=e^{-\beta E_n^0}/Z_0$ in Eq.\eqref{transprob}. 

We are interested in distinguishing fluctuations in an active equilibrium state from fluctuations in a Gibbs equilibrium state. The main difference is that the probability distribution of equilibrium work fluctuation is $p_w(x)\neq\delta(x)$ for the former, reflecting an active agent,  and $p_w(x)=\delta(x)$ for the latter, reflecting a passive agent.

To investigate other differences, we consider our charging map ${\mathcal E}$ and a thermal map ${\mathcal E}^{\rm Thm}$ for a qubit. The map ${\mathcal E}^{\rm Thm}$ is obtained by coupling the qubit to an auxiliary thermal qubit with $V=a(\sigma_S^+\sigma_B^-+\sigma_S^-\sigma_B^+)$, and tracing out the auxiliary system. 
The resulting map is thermal (i.e., a map with the Gibbs equilibrium state), and the transition matrix for this process is
\[
T^{\rm Thm}=\left(\begin{array}{cc}
1-\frac{e^{-\beta \frac{h}{2}}}{Z}g(a,0) & \frac{e^{\beta\frac{h}{2}}}{Z}g(a,0)\\
\frac{e^{-\beta\frac{h}{2}}}{Z}g(a,0) & 1-\frac{e^{\beta \frac{h}{2}}}{Z}g(a,0)
\end{array}\right)
\]
where $g(a,0)=\sin^2(\tau a/\hbar)$ and $Z=e^{\beta \frac{h}{2}}+e^{-\beta \frac{h}{2}}.$ $T^{\rm Thm}$ is a regular stochastic matrix if $g(a,0)\neq 0$. 
Te most crucial difference between $T^{\rm Thm}$ and $T_{1Q}$ in Eq.\eqref{Tqubitbattery} is the position of the factors $e^{\pm \beta h/2}$.

For the charging map, one can show
$P^{(L)}_{2\to 2}>P^{(L)}_{1\to 1}$, reflecting the higher population of the excited state in the active equilibrium.
Instead, for the thermal map $P^{(L){\rm Thm}}_{1\to 1}>P^{(L){\rm Thm}}_{2\to 2}$, reflecting the higher population of the ground state in Gibbs equilibrium. 
On the other hand, energy fluctuations due to $1\leftrightarrow 2$ transitions are qualitatively similar if $g(a,h)\approx g(a,0)$ for processes with finite $L$ but are indistinguishable for $L\to\infty$. Indeed,
for $L\to\infty$ we have
\[
P^{(\infty){\rm Thm}}_{1\to 2}=P^{(\infty){\rm Thm}}_{2\to 1}=\frac{1}{Z^2},\quad P^{(\infty){\rm Thm}}_{1\to 1}=\frac{e^{\beta h}}{Z^2},\quad 
P^{(\infty){\rm Thm}}_{2\to 2}=\frac{e^{-\beta h}}{Z^2}
\]
and for the charging map
\[
P^{(\infty)}_{2\to 1}=P^{(\infty)}_{1\to 2}=\frac{1}{Z^2},\quad
P^{(\infty)}_{2\to 2}=\frac{e^{\beta h}}{Z^2},\quad
P^{(\infty)}_{1\to 1}=\frac{e^{-\beta h}}{Z^2}.
\]

\subsection{Two-qubit battery}

We consider a two-qubit battery with Hamiltonian~\cite{BarraBattery}
\[
H_S=\frac{h}{2}\left(\sigma^z_1+\sigma^z_2\right)+J\left(\sigma^x_1\sigma^x_2+\sigma^y_1\sigma^y_2\right),
\]
coupled with
\[
V=J'(\sigma_{B}^x\sigma_1^x+\sigma_{B}^y\sigma_1^y),
\]
to auxiliary systems with Hamiltonian $H_B=\frac{h}{2}\sigma_{B}^z$, in the thermal state. 
The corresponding map ${\mathcal E}$ has the equilibrium state $\omega_{\beta}(H_0)$ with $H_0=\frac{h}{2}\left(\sigma^z_1+\sigma^z_2\right)$.

The eigenvalues and eigenvectors of $H_S$ and $H_0$ in the basis defined by $\sigma^z\ket{\uparrow}=\ket{\uparrow}$ and $\sigma^z\ket{\downarrow}=-\ket{\downarrow}$ are  
\begin{align}
E_3&=h,        & E^0_3&=h, &       \ket{3}&=\ket{\uparrow\uparrow},\\
E_4&=2J,      & E^0_4&=0, &       \ket{4}&=(\ket{\uparrow\downarrow}+\ket{\downarrow\uparrow})/\sqrt{2},\\
E_1&=-2J,     & E^0_1&=0, &      \ket{1}&=(\ket{\uparrow\downarrow}-\ket{\downarrow\uparrow})/\sqrt{2}, \\
E_2&=-h,       & E^0_2&=-h, &     \ket{2}&=\ket{\downarrow\downarrow}.
\end{align}

We take $2J>h>0$ such that $E_{i+1}>E_i$. The permutation that orders $E^0_{\pi_{i+1}}\geq E^0_{\pi_i}$ is $(\pi_1,\pi_2,\pi_3,\pi_4)=(2,1,4,3)$. 
Thus on the above basis, the equilibrium state is
\[
\omega_{\beta}(H_0)=\frac{e^{-\beta h} }{Z_0}\ket{3}\bra{3}+\frac{1}{Z_0}(\ket{1}\bra{1}+\ket{4}\bra{4})+\frac{e^{\beta h} }{Z_0}\ket{2}\bra{2},
\]
and the passive state for the system is 
\[
\sigma_{\omega_\beta(H_0)}=\frac{e^{\beta h} }{Z_0}\ket{1}\bra{1}+\frac{1}{Z_0}(\ket{2}\bra{2}+\ket{3}\bra{3})+\frac{e^{-\beta h} }{Z_0}\ket{4}\bra{4},
\]
where $Z_0=2+2\cosh(\beta h).$
The ergotropy of the equilibrium state ${\mathcal W}=\Tr[H_S(\omega_{\beta}(H_0)-\sigma_{\omega_\beta(H_0)})]$ is
\[
{\mathcal W}=(2J-h)\frac{\sinh \beta h}{1+\cosh \beta h}.
\]
The work done in the charging process $\sigma_{\omega_\beta(H_0)}\to \omega_\beta(H_0)$ is
\[
W_R=2J \frac{\sinh \beta h}{1+\cosh \beta h}.
\]
We see that the thermodynamic efficiency is $\eta_{\rm th} = \mathcal W/W_R=1-\frac{h}{2J}$ independently of the inverse temperature $\beta$.

The recharging process in this two-qubit battery (2Q) is determined by the stochastic matrix (see Eq.\eqref{transprob})
\begin{equation}
\label{T4x4}
T_{2Q}=\frac{1}{(J^2+J'^2)^2}\left(
\begin{array}{cccc}
\Phi^2 & \frac{2}{(1+e^{\beta h})}\Phi \Psi & \frac{2e^{\beta h}}{(1+e^{\beta h})}\Phi \Psi & \Psi^2\\
\frac{2e^{\beta h}}{(1+e^{\beta h})}\Phi \Psi & \frac{e^{\beta h}(J^2+J'^2)^2+\Delta}{(1+e^{\beta h})} & 0 & \frac{2e^{\beta h}}{(1+e^{\beta h})}\Phi \Psi\\
\frac{2}{(1+e^{\beta h})}\Phi \Psi& 0 & \frac{(J^2+J'^2)^2+e^{\beta h}\Delta}{(1+e^{\beta h})} & \frac{2}{(1+e^{\beta h})}\Phi \Psi\\
\Psi^2 & \frac{2}{(1+e^{\beta h})}\Phi \Psi & \frac{2e^{\beta h}}{(1+e^{\beta h})}\Phi \Psi & \Phi^2
\end{array}\right),
\end{equation}
with
\[
\Phi=J^2+J'^2\cos^2(\frac{\tau}{\hbar}\sqrt{J^2+J'^2}),
\quad
\Psi=J'^2\sin^2(\frac{\tau}{\hbar}\sqrt{J^2+J'^2}),
\quad
\Delta=(\Phi-\Psi)^2,
\]
which is a regular stochastic matrix excepts at points with $\Psi=0$ or $\Phi=0$, as one can check by computing $T^2$. 

\subsubsection{fluctuating efficiency}

For the fluctuating efficiency Eq.\eqref{flucteff2} we have 
\begin{align}
\eta_{12}&=\eta_{13}=\eta_{21}=\eta_{34}=\eta_{42}=\eta_{43}=1-\frac{h}{2J}\\
\eta_{14}&=\eta_{41}=\frac{1}{2}(1-\frac{h}{2J})\\
\eta_{23}&=\eta_{32}=\infty\\
\eta_{24}&=\eta_{31}=-(1-\frac{h}{2J})\\
\end{align}
and all $\eta_{nn}=\infty$.
Its distribution follows from Eq.\eqref{etadistrinm} and it is
\[
p_\eta(x)=\delta(x-\infty)\mathrm{P}_\infty+\delta\left(x-1+\frac{h}{2J}\right)\mathrm{P}_{(1-\frac{h}{2J})}+\delta\left(x+1-\frac{h}{2J}\right)\mathrm{P}_{-(1-\frac{h}{2J})}+\delta\left(x-\frac{1}{2}+\frac{h}{4J}\right)\mathrm{P}_{(1/2)(1-\frac{h}{2J})}\,
\]
with
\begin{align}
&\mathrm{P}_\infty=P_{3\to2}+P_{2\to3}+\sum_n P_{n\to n}=3\frac{(e^{\beta h}+e^{-\beta h})}{Z_0^2},\label{28}\\
&\mathrm{P}_{(1-\frac{h}{2J})}=P_{1\to2}+P_{1\to3}+P_{2\to1}+P_{3\to4}+P_{4\to2}+P_{4\to3}=\frac{2+(e^{\beta h}+e^{-\beta h})^2}{Z_0^2},\label{29}\\
&\mathrm{P}_{-(1-\frac{h}{2J})}=P_{3\to1}+P_{2\to4}=\frac{2}{Z_0^2},\label{30}\\
&\mathrm{P}_{(1/2)(1-\frac{h}{2J})}=P_{1\to4}+P_{4\to1}=\frac{(e^{-\beta h}+e^{\beta h})}{Z_0^2}\label{31}.
\end{align}
The explicit formulas at the right are valid for parameters $\tau,J$ and $J'$ in which $T_{2Q}$ is regular.

\begin{figure}[H] 
\centering
  \includegraphics[width=0.4\textwidth]{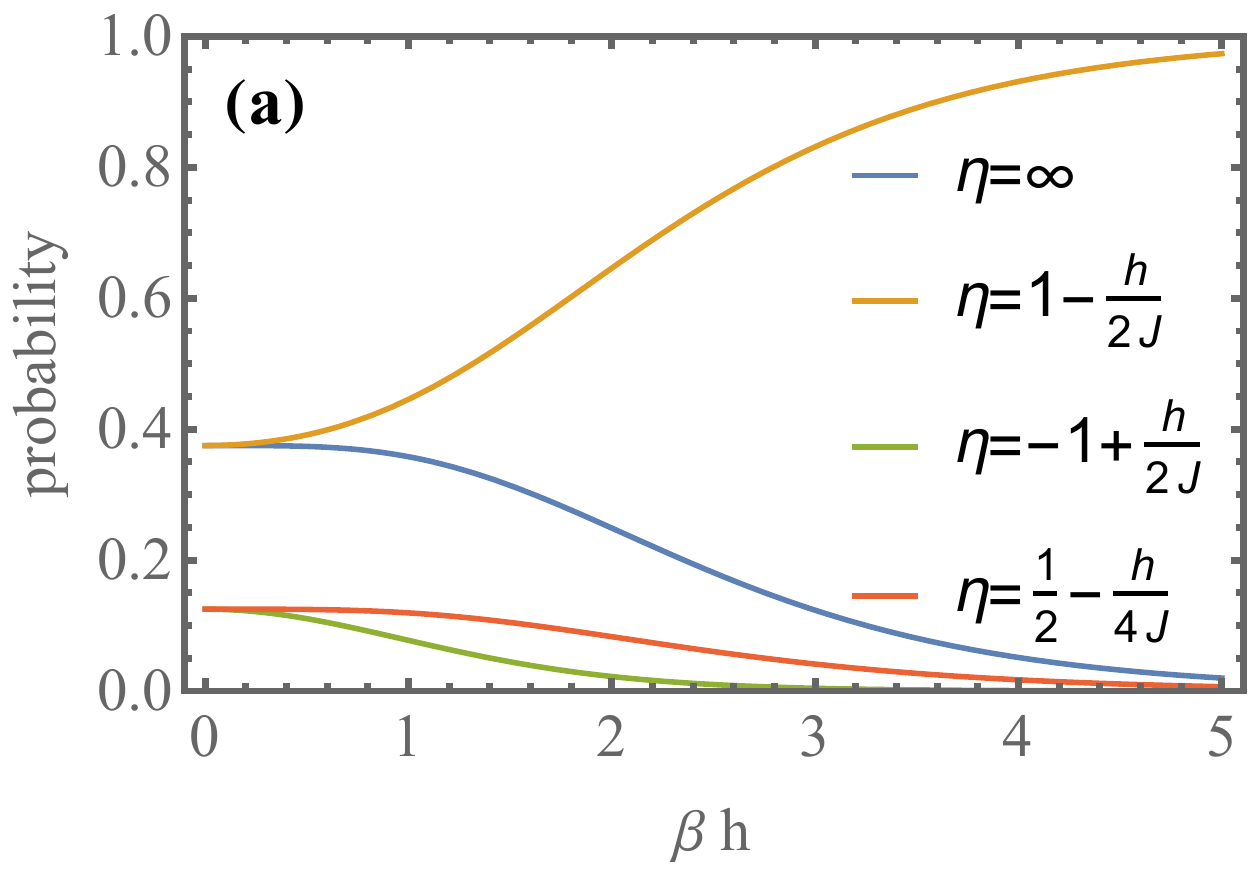}
   \includegraphics[width=0.4\textwidth]{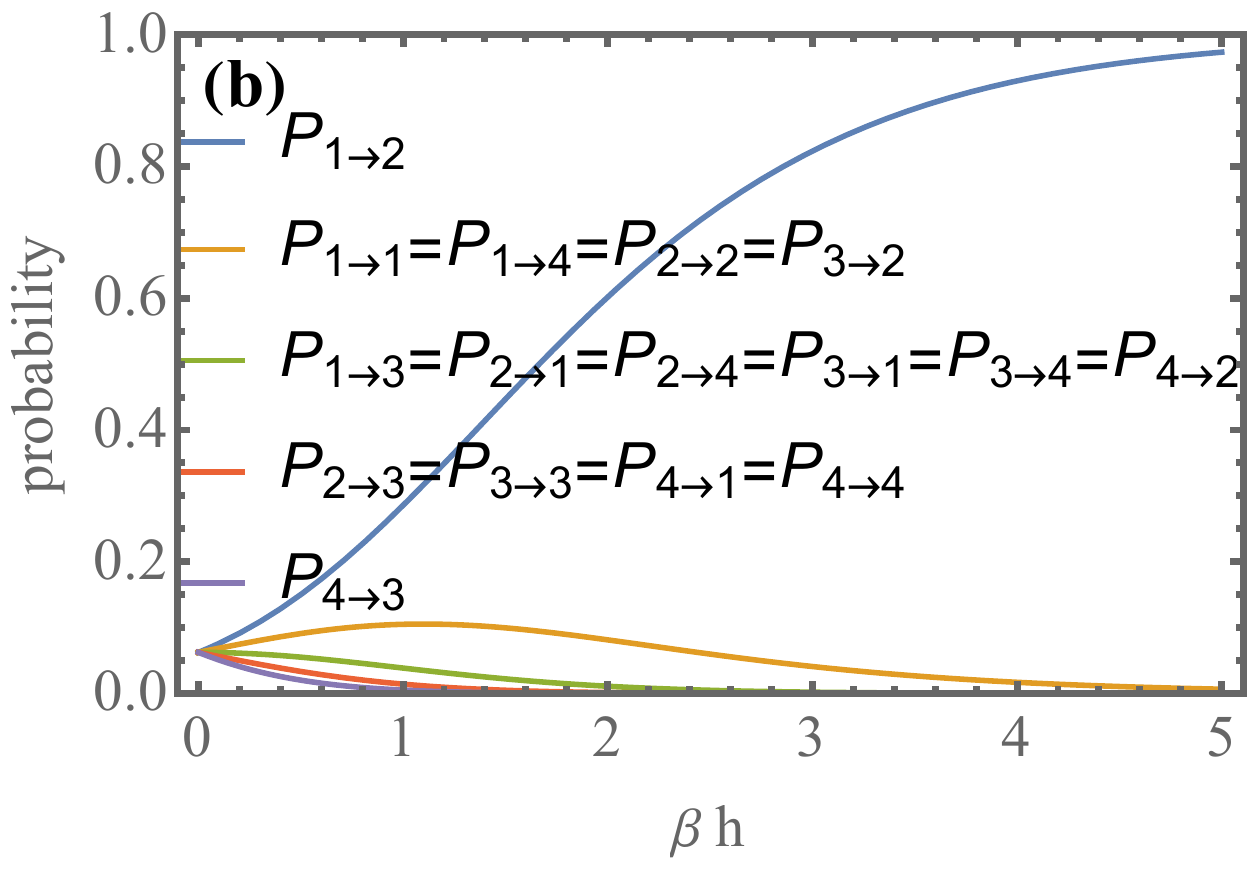}
    \caption{For the 2-qubit battery: (a) Plots of $\mathrm{P}_\eta$ as function of $\beta h$  and (b) Plots of $P_{n\to m}$ given by Eq.\eqref{flucteff2} for the two-qubit battery.}
    \label{fig:Thermo}
\end{figure}

In Fig.~\ref{fig:Thermo}a, we plot the probabilities $\mathrm{P}_\eta$ in Eqs.~\eqref{28}, \eqref{29}, \eqref{30}, and \eqref{31} as a function of $\beta h$. We see that for small $\beta h$, where many transitions are assisted by heat, the average efficiency does not exist. On the other hand, when $\beta h\gg 1$, the efficiency goes to the thermodynamic efficiency with probability one because the work becomes deterministic. In Fig.~\ref{fig:Thermo}b, we see that $P_{1\to 2}$ goes to one in that limit. The second in importance are the $P_{1\to 4}$ associated with the largest charge but not very efficient transition and the $P_{3\to 2}$ contributing to $\eta=\infty$.


\subsubsection{Heat and work fluctuations in the partial recharging process}

Here, we consider the process ${\mathcal E}^L$ starting in the state $\sigma_{\omega_\beta(H_0)}$ and evaluate the heat and work distributions. Hence, we consider Eq.\eqref{ecc:dist trabajo eq. map}
and Eq.\eqref{ecc:dist calor eq. map} with $P_{n \to m}^{(L)}=(T^L)_{m|n}\frac{e^{-\beta E_{\pi_n}^0}}{Z0}$, with the permutation $\pi$ ordering the eigenvalues of $H_0$ increasingly. 

For the two-qubit battery we obtain
\begin{align}
p_w^{(L)}(x)&=\delta(x)A^{(L)}_0+\delta(x-4J)A^{(L)}_1+\delta(x-2J)A^{(L)}_3+\delta(x+4J)A^{(L)}_2+\delta(x+2J)A^{(L)}_4,\\
p_q^{(L)}(x)&=\delta(x)B^{(L)}_0+\delta(x-h)B^{(L)}_4+\delta(x-2h)B^{(L)}_2+\delta(x+h)B^{(L)}_3+\delta(x+2h)B^{(L)}_1
\end{align}
with

\begin{align}
A^{(L)}_0&=P^{(L)}_{2\to 3}+P^{(L)}_{3\to 2},& B^{(L)}_0&=P^{(L)}_{1\to 4}+P^{(L)}_{4\to 1},\\
A^{(L)}_1&=P^{(L)}_{1\to 4},& B^{(L)}_1&=P^{(L)}_{3\to 2}, \\
A^{(L)}_2&=P^{(L)}_{4\to 1},& B^{(L)}_2&=P^{(L)}_{2\to 3},\\
A^{(L)}_3&=P^{(L)}_{1\to 2}+P^{(L)}_{1\to 3}+P^{(L)}_{2\to 4}+P^{(L)}_{3\to 4},& B^{(L)}_3&=P^{(L)}_{1\to 2}+P^{(L)}_{3\to 1}+P^{(L)}_{4\to 2}+P^{(L)}_{3\to 4},\\
A^{(L)}_4&=P^{(L)}_{2\to 1}+P^{(L)}_{3\to 1}+P^{(L)}_{4\to 2}+P^{(L)}_{4\to 3},& B^{(L)}_4&=P^{(L)}_{2\to 1}+P^{(L)}_{1\to 3}+P^{(L)}_{2\to 4}+P^{(L)}_{4\to 3},\\
\end{align}
where $A_i^{(L)}\neq B_i^{(L)}$ for finite $L$ but $A_i^{(\infty)}= B_i^{(\infty)}$ with 
\[
A_0^{(\infty)}=
\frac{6 \cosh\beta h}{Z_0^2},\quad
A_1^{(\infty)}=\frac{e^{\beta h}}{Z_0^2},\quad A_2^{(\infty)}=\frac{e^{-\beta h}}{Z_0^2},\quad
A_3^{(\infty)}=\frac{e^{2\beta h}+3}{Z_0^2}, \quad A_4^{(\infty)}=\frac{3+e^{-2\beta h}}{Z_0^2}.
\]
This means that the average work $W^{(L)}$ and average heat $Q^{(L)}$
\[
W^{(L)}=
2J(A^{(L)}_3-A^{(L)}_4)+4J(A^{(L)}_1-A^{(L)}_2)\xrightarrow[L\to\infty]{}\frac{2J\sinh\beta h}{1+\cosh\beta h}
\]
\[
Q^{(L)}=h(B^{(L)}_4-B^{(L)}_3)+2h(B^{(L)}_2-B^{(L)}_1)\xrightarrow[L\to\infty]{}\frac{-h \sinh\beta h}{1+\cosh\beta h}
\]
becomes proportional when $L\to\infty$. 

\begin{figure}[H] 
\centering
\begin{tabular}{ccc}
  \includegraphics[width=0.3\textwidth]{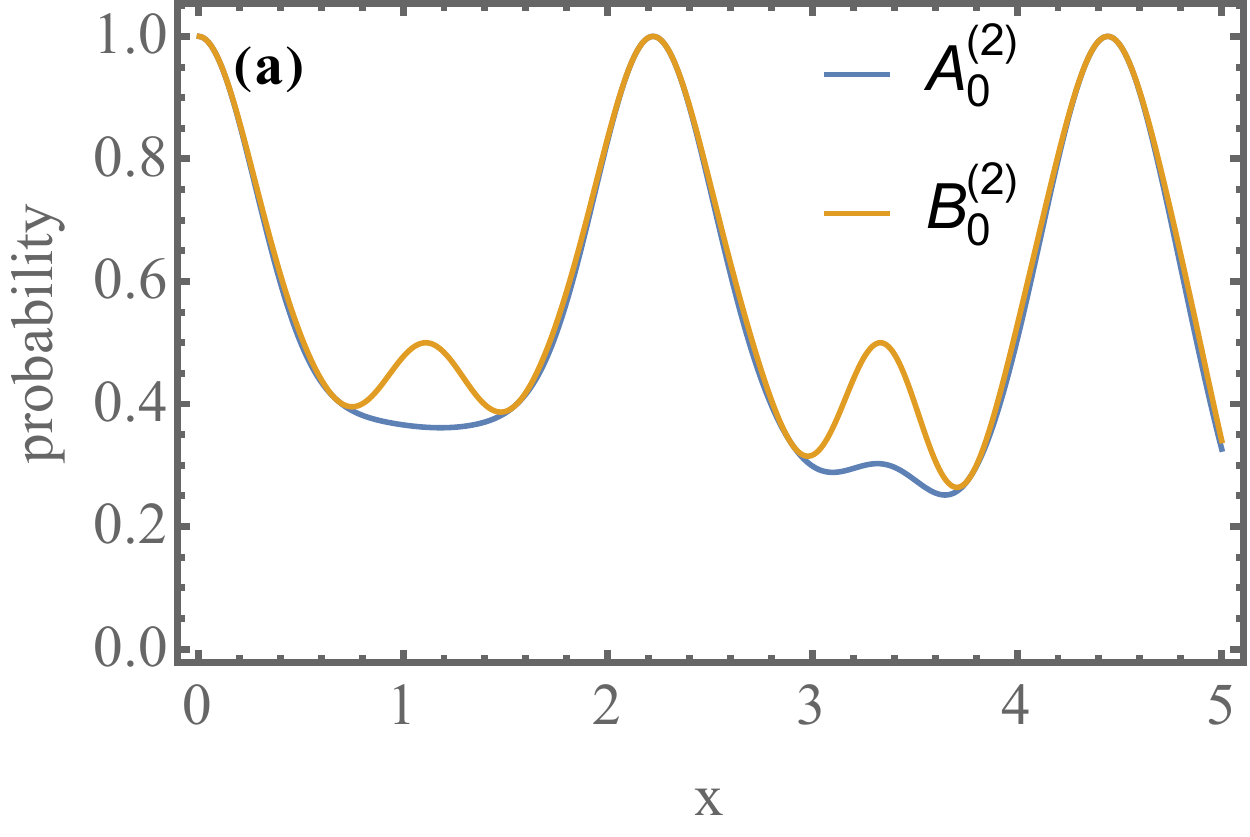}&
    \includegraphics[width=0.3\textwidth]{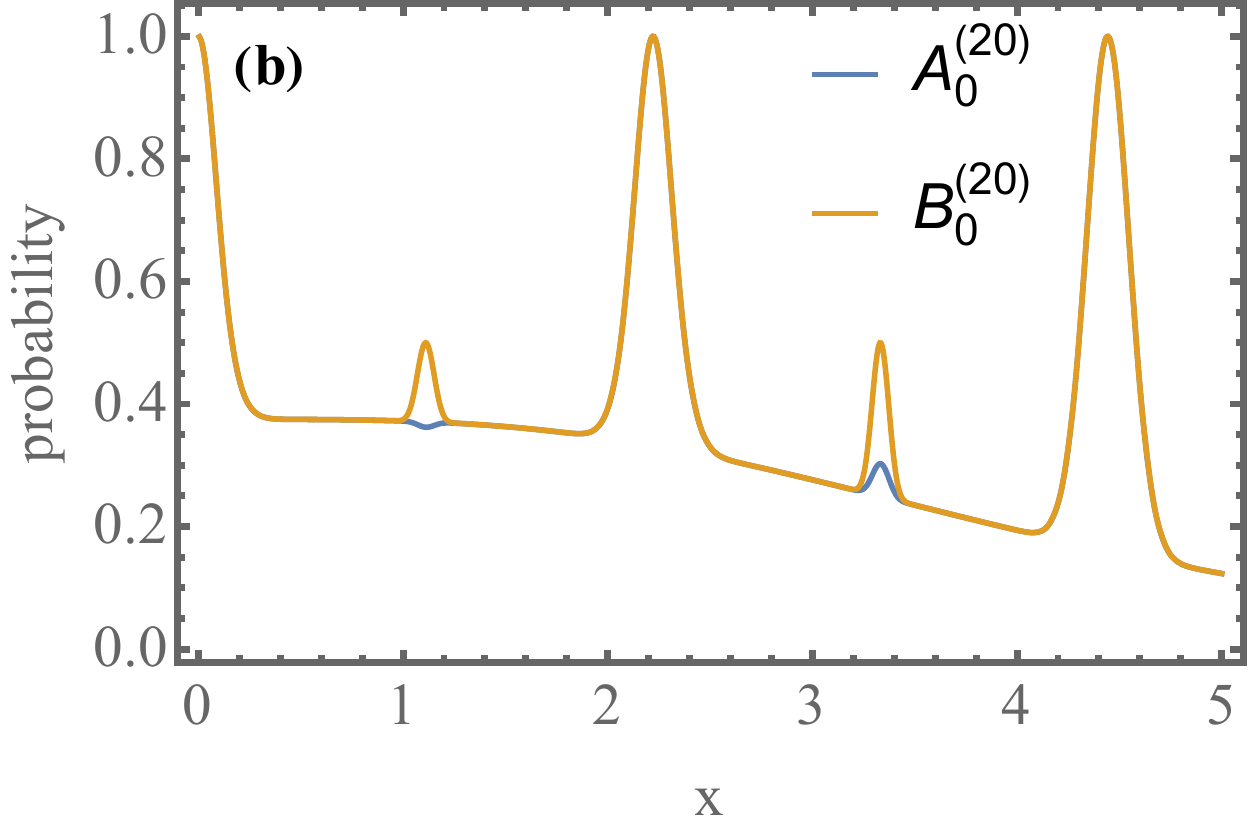}&
     \includegraphics[width=0.3\textwidth]{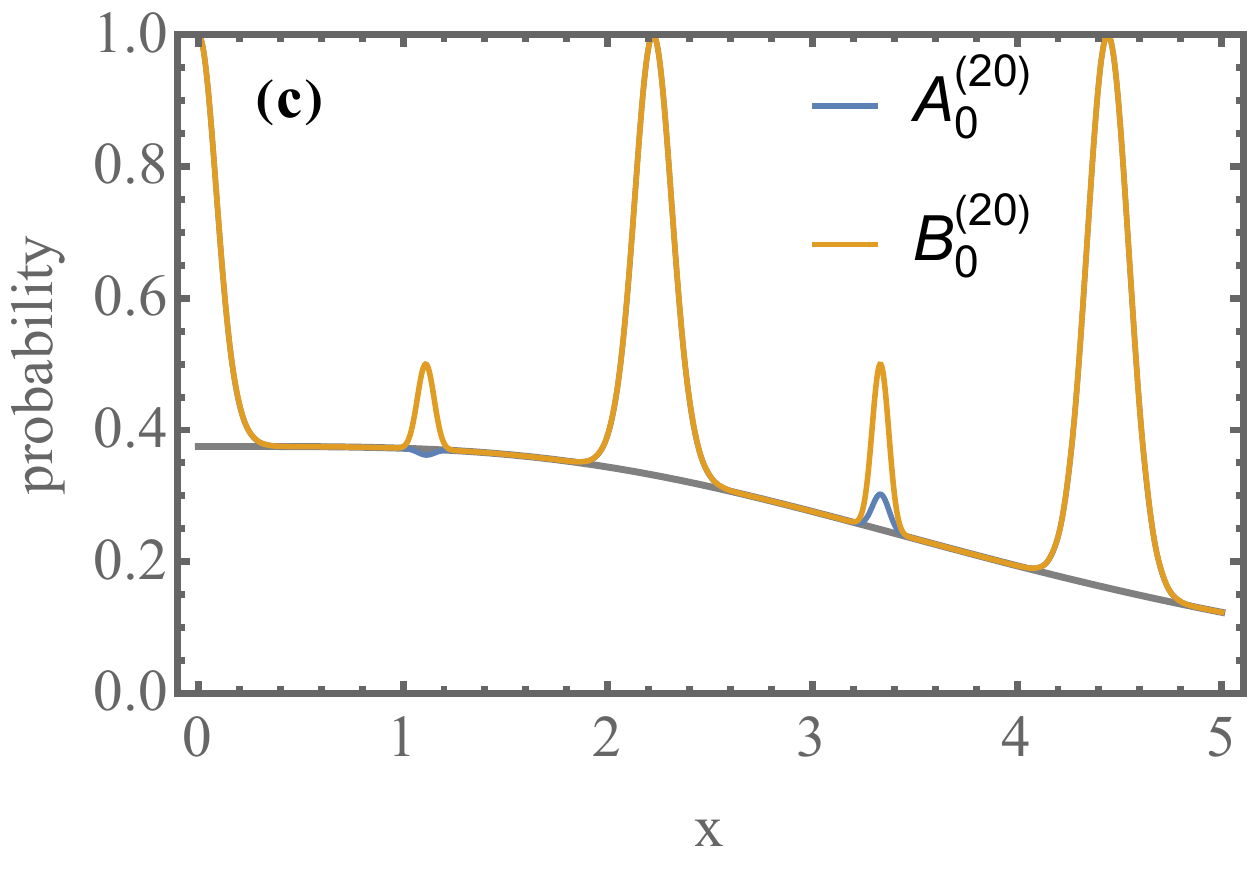}\\
    \includegraphics[width=0.3\textwidth]{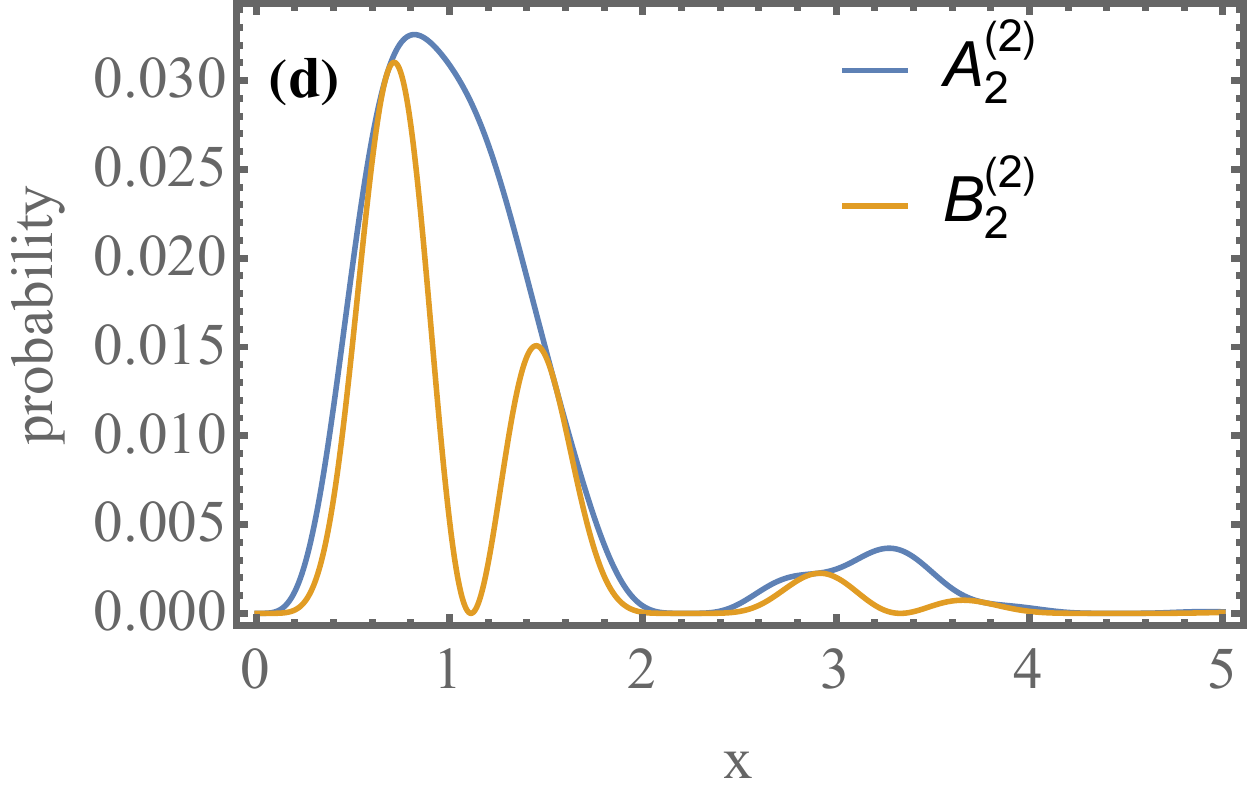}   &
      \includegraphics[width=0.3\textwidth]{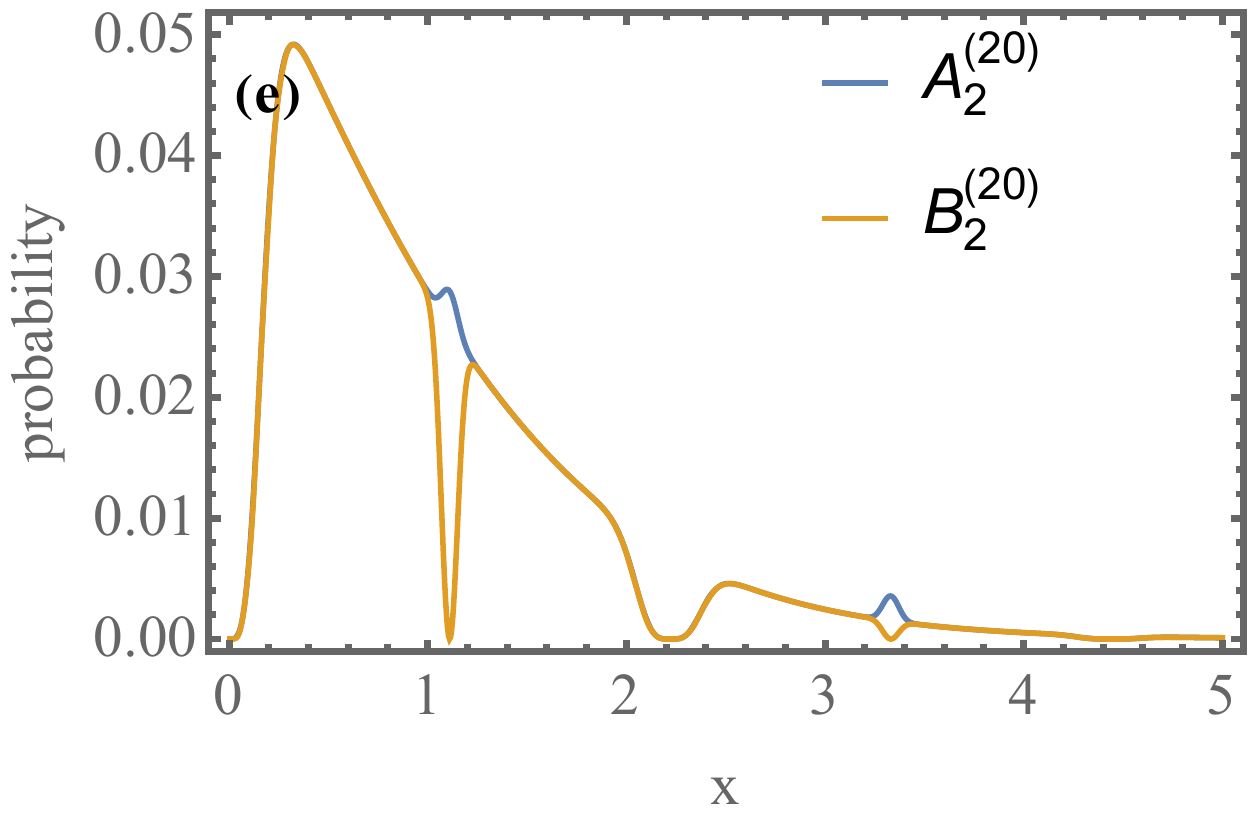}&
     \includegraphics[width=0.3\textwidth]{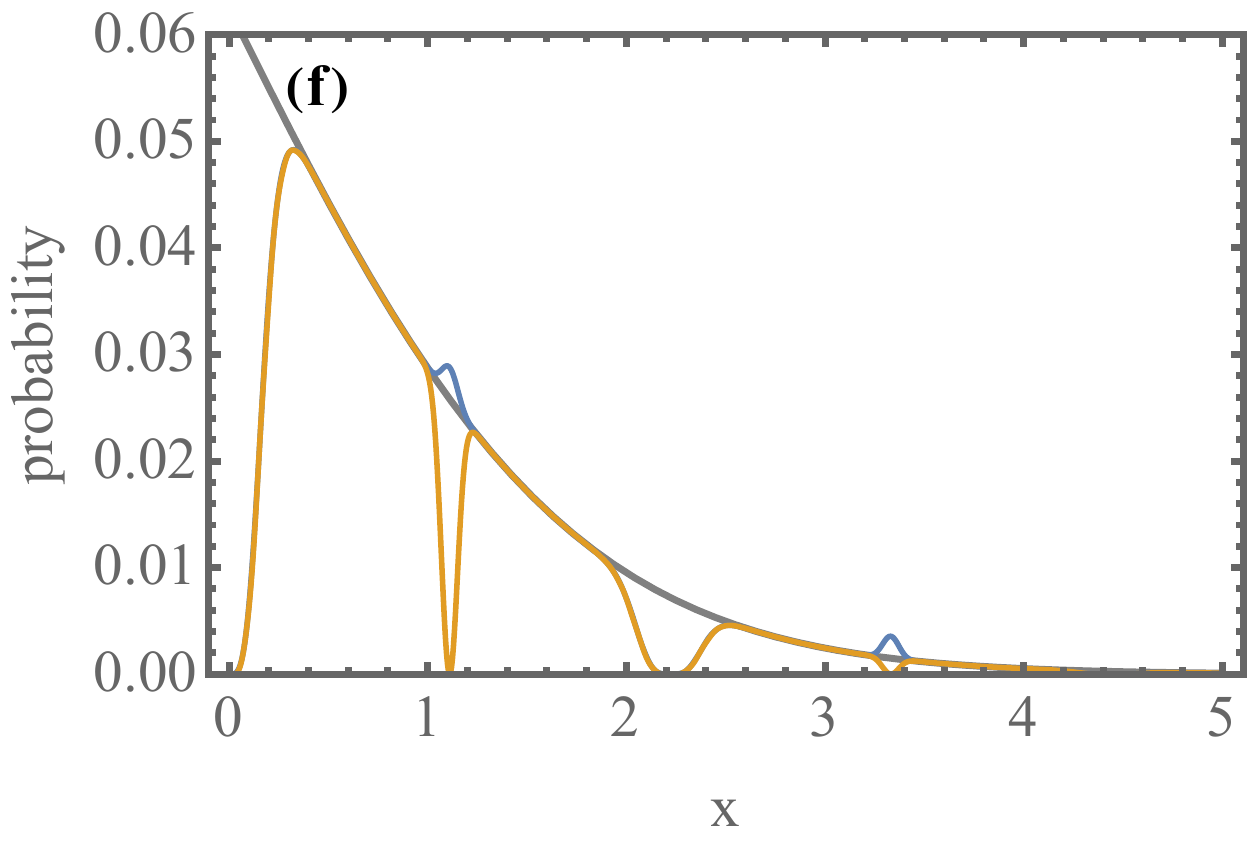}\\
   \end{tabular}  
    \caption{ Plots of the probabilities $A^{(L)}_i$ and $B^{(L)}_i$, with $L=2$ at the left (a and d) and $L=20$ at the center (b and e) with $i=0$ at the top (a and b) and $i=2$ at the bottom (d and e). At the right (c and f) we superpose the analytical result $A^{(\infty)}_i$ and $B^{(\infty)}_i$ to the data at the center for $L=20$. For the numerical computation we take $\beta=\tau/\hbar=1,J=J'=x$ and $h=0.6 x$. }
    \label{fig:11}
\end{figure}

Since Markov chains converge exponentially to the stationary state, it is unnecessary to consider a large $L$ to observe the asymptotic distribution. However, since the convergence rate  depends on the map's parameters, we see deviations from it near the points where $\Phi=0$ or $\Psi=0$ in Eq.\eqref{T4x4}.  To illustrate this, we plot in Figure~\ref{fig:11} the probabilities $A^{(L)}_0,B^{(L)}_0,A^{(L)}_2$ and $B^{(L)}_2$ for various values of $L$ and varying a map parameter.

\section{Discussion}
\label{secCONC}

We have studied stochastic fluctuations in repeated interaction processes subjected to the two-point energy measurement scheme. Because map ${\mathcal E}$ has an equilibrium state, all quantities are expressed in terms of system properties simplifying their study because one does not require measuring the environment. 
We have shown that the equilibrium distribution of the map dominates the distributions, except at particular points in the parameter space of the map, where its details become essential. Near these zones, the convergence rate towards the asymptotic value is low, requiring larger values of $L$ to reach it. The quantum aspect of the system is relevant near these zones since the Planck constant appears in the parameters that set the convergence rate to the stationary state. 
We have applied these results to study active equilibrium fluctuations, fluctuations in the charging process of a quantum battery, and efficiency fluctuations of the cycle charging and extracting energy for the battery in two examples. The fluctuating efficiency converges to the thermodynamic efficiency of these examples in the low-temperature limit, where work fluctuations are negligible. On the other hand, at large temperatures, where heat assists many transitions, the efficiency may become infinite, preventing the existence of the average. 

For future research, it would be interesting to extend the results obtained here for the single-cycle efficiency to the case of an arbitrary number of cycles. As this number increases, universal statistical behaviors have been shown to appear in other machines~\cite{ef1,ef2,ef11}.

\section*{Acknowledgments}

F.B. gratefully acknowledges the financial support of FONDECYT grant 1191441 and the Millennium Nucleus “Physics of active matter” of ANID (Chile).

\section*{Appendices}
\begin{appendices}
\section{Distributions for maps with equilibrium} 
\label{sec:appendix}

Let us justify Eq.\eqref{psm1} and Eq.~\eqref{ecc:dist calor eq. map}.  Eq.~\eqref{ecc:dist energ eq. map} and Eq.~\eqref{ecc:dist trabajo eq. map} follow from the same argument. 

We can consider that the system $S$ and all the copies of system $B$ start uncorrelated in a product state. We measure the energy of that system and project the state to $\ket{i_1\cdots i_L n}$ with probability $\frac{e^{-\beta \sum_{k=1}^L\varepsilon_{i_k}}}{Z_B^L}p_{\rm ini}(n)$ because the copies of $B$ are in the Gibbs state. Then, the full system evolves unitarily by composing the unitary evolutions where at each time only the system $S$ with a copy $i$ of $B$ is interacting. This is represented by the product $U_L\cdots U_1$, and the global state is $U_L\cdots U_1\ket{i_1\cdots i_L n}$. Then we measure the energy of $S$ and of each copy of $B$. According to the Born rule after the measurement the total system is is the state $\ket{j_1\cdots j_L m}$ with probability
\begin{equation}
P_\gamma^{(L)}=|\langle j_1\cdots j_L n_{a_L}|U_L\cdots U_1|i_1\cdots i_L n_{a_0}\rangle|^2 \frac{e^{-\beta \sum_{n=1}^L\varepsilon_{i_n}}}{Z_b^L}p_i(n_{a_0}).
\end{equation}
More details are found in~\cite{StochPRE}. 

We use that result to derive Eq.~\eqref{ecc:dist calor eq. map} and by extension, all other distributions for maps with equilibrium. Consider that
\[
\langle j_1\cdots j_L n_{a_L}|U_L\cdots U_1|i_1\cdots i_Ln_{a_0}\rangle=
\sum_{a_1a_2..a_{L-1}}\langle n_{a_L}j_L|U_L|n_{a_{L-1}}i_L\rangle\cdots   \langle n_{a_2} j_2|U_2|n_{a_1}i_2\rangle\langle n_{a_1} j_1|U_1|n_{a_0}i_1\rangle
\]
Because $[H_0+H_B,U_k]=0$, the generic transition 
$\langle n_{a_k}j_{k}|U_k|n_{a_{k-1}}i_{k}\rangle=0$ unless $E^0_{a_k}+\varepsilon_{j_k}=E^0_{a_{k-1}}+\varepsilon_{i_k}$. Thus in every trajectory $\gamma$ with non-vanishing probability we have
\[
q_\gamma=\sum_k (\varepsilon_{i_k}-\varepsilon_{j_k})=\sum_k(E^0_{a_k}-E^0_{a_{k-1}})=E^0_{a_L}-E^0_{a_1}.
\]
Hence
\[
p_q^{(L)}(x)=\sum_\gamma\delta(x-q_\gamma)P^{(L)}_\gamma=\sum_\gamma\delta(q-(E^0_{a_L}-E^0_{a_0}))P^{(L)}_\gamma=\sum_{a_L,a_0}\delta(q-(E^0_{a_L}-E^0_{a_0}))\sum_{\gamma:a_L,a_0}P^{(L)}_\gamma,
\]
where in the last sum, we add over all trajectories $\gamma$ starting at $n_{a_0}$ and ending at $n_{a_L}$. 
This correspond to taking the traces over all systems $B$ that interacted with $S$ and thus $\sum_{\gamma:a_L,a_0}P^{(L)}_\gamma=\bra{n_{a_L}}{\mathcal E}^L(\ket{n_{a_0}}\bra{n_{a_0}})\ket{n_{a_L}}p_i(n_{a_0})$.

\end{appendices}


\begin{thebibliography}{999}

\bibitem{rep-int} 
Attal, S.; Pautrat, Y.  From repeated to continuous quantum interactions. {\em Ann. Inst. Henri Poincar\'{e}} {\bf 2006}, {\em 7}, 59. 
\bibitem{Attal2}  Attal, S.; Joye,  A. Weak coupling and continuous limits for repeated quantum interactions. {\em J. Stat. Phys.} {\bf 2007}, {\em 126}, 1241.  
\bibitem{rep-int2}  Giovannetti, V.; Palma, G. M. Master equations for correlated quantum channels. {\em Phys. Rev. Lett. }{\bf 2012}, {\em 108}, 040401.
\bibitem{Karevski} Karevski, D.; Platini, T. Quantum nonequilibrium steady states induced by repeated interactions. {\em Phys. Rev. Lett.} {\bf 2009}, {\em 102}, 207207. 
\bibitem{rep-int3} Lorenzo, S.; Ciccarello, F.; Palma, G. M.; Vacchini, B. Quantum Non-Markovian Piecewise Dynamics from Collision Models. {\em Open Systems \& Information Dynamics} {\bf 2017}, {\em 24}, 1740011.
\bibitem{rep-int4} Strasberg, P. Repeated interactions and quantum stochastic thermodynamics at strong coupling {\em Phys. Rev. Lett.} {\bf 2019}, {\em 123}, 180604.

\bibitem{maser1} Cresser, J. D. Quantum-field model of the injected atomic beam in the micromaser. {\em Phys. Rev. A} {\bf 1992}, {\em 46}, 5913.
\bibitem{maser2} Englert, B.G.; Morigi, G. Five lectures on dissipative master equations. In Coherent evolution in noisy environments—Lecture Notes in Physics; Buchleitner, A., Hornberger, K., Eds.; Springer: Berlin/Heidelberg, Germany, 2002; p. 611.
\bibitem{maser3} Walther, H. The Deterministic generation of photons by cavity quantum electrodynamics, Chapter 1 of Elements of Quantum Information, (Wiley, 2007). 
\bibitem{maser4} Ciccarello, F. Collision models in quantum optics. {\em Quantum Measurements and Quantum Metrology}  {\bf 2017}, {\em 4}, 53.
\bibitem{Qth1} Kosloff, R. Quantum thermodynamics: A dynamical viewpoint. {\em Entropy} {\bf 2013}, {\em 15}, 2100. 
\bibitem{Qth2}Kosloff, R.;  Levy, A. Quantum heat engines and refrigerators: Continuous devices. {\em Annu. Rev. Phys. Chem.} {\bf 2014}, {\em 65}, 365. 
\bibitem{Qth3}  Vinjanampathy, S.;  Anders, J. Quantum thermodynamics. {\em Contemp. Phys.} {\bf 2016}, {\em 57}, 545.
\bibitem{Qth4} Goold, J.; Huber, M.; Riera, A.; del Rio, L.; Skrzypczyk, P. The role of quantum information in thermodynamics: a topical review.  {\em J. Phys. A: Math. Theor.} {\bf 2016}, {\em 49}, 143001.
\bibitem{Qth5} Strasberg, P. {\em Quantum stochastic thermodynamics: Foundations and selected applications}. Oxford University Press: Oxford, UK, 2022.

\bibitem{info1} Deffner, S.; Christopher J. Information processing and the second law of thermodynamics: An inclusive, Hamiltonian approach.  {\em Phys. Rev. X} {\bf 2013}, {\em 3}, 041003.
\bibitem{info2} Strasberg, P.; Schaller, G.; Brandes, T.; Esposito, M. Thermodynamics of a physical model implementing a Maxwell demon. {\em Phys. Rev. Lett. } {\bf 2013}, {\em 110}, 040601.
\bibitem{info3} Landi, G. T. Battery charging in collision models with Bayesian risk strategies. {\em Entropy} {\bf 2021}, {\em 23}, 1627.
\bibitem{Esposito rep. int.} Strasberg, P.; Schaller, G.; Brandes, T.; Esposito M. Quantum and information thermodynamics: A unifying framework based on repeated interactions. {\em Phys. Rev. X} {\bf 2017}, {\em 7}, 021003. 

\bibitem{rep-int-engine} Molitor, O. A. D.;  Landi, G. T. Stroboscopic two-stroke quantum heat engines. {\em Phys. Rev. A} {\bf 2020}, {\em 102}, 042217. 
\bibitem{DenzlerLutz} Denzler, T.; Lutz, E. Efficiency fluctuations of a quantum heat engine. {\em Phys. Rev. Res.} {\bf 2020} ,{\em 2},  032062(R).

\bibitem{enginePhil}Strasberg, P.; W\"achtler, C.W.; Schaller, G. Autonomous Implementation of Thermodynamic Cycles at the Nanoscale. {\em Phys. Rev. Lett.} {\bf 2021}, {\em 126}, 180605.

\bibitem{PreB} Purkayastha, A.; Guarnieri, G.; Campbell, S.; Prior, J.; Goold, J. Periodically refreshed quantum thermal machines {\em arXiv} {\bf 2202}, arXiv:2202.05264.

\bibitem{QBRepInt} Seah, S.; Perarnau-Llobet, M.; Haack, G.; Brunner, N.;  Nimmrichter, S. Quantum speed-up in collisional battery charging. {\em Phys. Rev. Lett.} {\bf 2021}, {\em 127}, 100601.  

\bibitem{QBRepInt2} Shaghaghi, V.; Palma, G. M.; Benenti, G. Extracting work from random collisions: A model of a quantum heat engine. {\em Phys. Rev. E}  {\bf 2022}, {\em 105}, 034101.
\bibitem{BarraBattery} Barra, F. Dissipative charging of a quantum battery. {\em Phys. Rev. Lett.}  {\bf 2019}, {\em 122}, 210601. 
\bibitem{firstQB} Alicki, R.; Fannes, M. Entanglement boost for extractable work from ensembles of quantum batteries. {\em Phys. Rev. E} {\bf 2013}, {\em 87}, 042123. 
\bibitem{quantacell1}Binder,  F. C.; Vinjanampathy, S.; Modi, K; Goold, J. Quantacell: powerful charging of quantum batteries. {\em New J. Phys.}  {\bf 2015}, {\em 17}, 075015. 
\bibitem{quantacell2} Campaioli, F.; Pollock, F. A.; Binder, F. C.; C\'eleri, L.; Goold, J.; Vinjanampathy, S.; Modi, K. Enhancing the charging power of quantum batteries. {\em Phys. Rev. Lett.} {\bf 2017}, {\em 118}, 150601.
\bibitem{campisibatt} Ferraro, D.; Campisi, M.; Andolina, G. M.; Pellegrini, V.;  Polini, M. High-power collective charging of a solid-state quantum battery. {\em Phys. Rev. Lett.} {\bf 2018}, {\em 120}, 117702.
\bibitem{qb1} Hovhannisyan, K.; Barra, F.; Imparato, A. Charging assisted by thermalization. {\em Phys. Rev. Res.} {\bf 2020}, {\em 2} 033413.
\bibitem{qb2} Barra, F.; Hovhannisyan, K.; Imparato, A. Quantum batteries at the verge of a phase transition. {\em New. J. Phys.} {\bf 2022}, {\em 24}, 015003.
\bibitem{qb3} Carrasco, J.; Hermann, C.; Maze, J.; Barra, F. Collective enhancement in dissipative quantum batteries. {\em arXiv} {\bf 2021}, arXiv:2110.15490.


\bibitem{PreB2} Purkayastha, A.; Guarnieri, G.; Campbell, S.; Prior, J.; Goold, J. Periodically refreshed baths to simulate open quantum many-body dynamics. {\em Phys. Rev. B} {\bf 2021}, {\em 104}, 045417.

\bibitem{rev1} Ciccarello, F.; Lorenzo, S.; Giovannetti, V.;  Palma, G. M. Quantum collision models: Open system dynamics from repeated interactions. {\em Phys. Rep. } {\bf 2022}, {\em 954}, 1.

\bibitem{rev2}  Campbell, S.; Vacchini, B. Collision models in open system dynamics: A versatile tool for deeper insights? {\em EPL} {\bf 2021}, {\em 133}, 60001.

\bibitem{BreuerBook}  Breuer, H.-P. and Petruccione, F., {\em The Theory of Open Quantum Systems}; Oxford University Press: Oxford, UK, 2002.


\bibitem{MHPPRE2015} Manzano, G.;  Horowitz, J. M.; Parrondo, J. M. R. Nonequilibrium potential and fluctuation theorems for quantum maps. {\em Phys. Rev. E} {\bf 2015}, {\em 92}, 032129.
\bibitem{HPNJP2013} Horowitz, J. M; Parrondo, J. M. R. Entropy production along non-equilibrium quantum jump trajectories. {\em  New J. Phys. } {\bf 2013}, {\em 15}, 085028.
\bibitem{MHPPRX2018} Manzano, G.; Horowitz, J. M.; Parrondo, J. M. R. Quantum fluctuation theorems for arbitrary environments:
Adiabatic and Nonadiabatic Entropy Production.  {\em Phys. Rev. X } {\bf 2018}, {\em 8} 031037. 
\bibitem{Campisi} Campisi, M.; H\"anggi, P.;  Talkner, P. Colloquium: Quantum fluctuation relations: Foundations and applications. {\em Rev. Mod. Phys.} {\bf 2011} , {\em 83}, 771. 

\bibitem{ef1} Verley, G.;  Esposito, M.; Willaert, T.; Van den Broeck, C. The unlikely Carnot efficiency.  {\em Nat. Commun.} {\bf 2014}, {\em 5}, 4721.
\bibitem{ef2}Verley, G.; Willaert, T.; Van den Broeck, C.; Esposito, M. Universal theory of efficiency fluctuations. {\em Phys. Rev. E} {\bf 2014}, {\em 90}, 052145.
\bibitem{ef3} Gingrich, T. R.;  Rotskoff, G. M.; Vaikuntanathan, S; Geissler, P. L. Efficiency and large deviations in time-asymmetric stochastic heat engines. {\em New J. Phys.} {\bf 2014}, {\em 16}, 102003.
\bibitem{ef4}Polettini, M.; Verley, G.; Esposito, M. Efficiency statistics at all times: Carnot limit at finite power. {\em Phys. Rev. Lett.} {\bf 2015}, {\em 114}, 050601.
\bibitem{ef5}Proesmans, K.; Cleuren, B.; ; Van den Broeck, C. Stochastic efficiency for effusion as a thermal engine. {\em Europhys. Lett.} {\bf 2015}, {\em 109}, 20004.
\bibitem{ef6}Proesmans, K.; Van den Broeck, C. Stochastic efficiency: Five case studies. {\em New J. Phys.} {\bf 2015}, {\em 17}, 065004.
\bibitem{ef7}Proesmans, K.; Dreher, Y.; Gavrilov, M.; Bechhoefer, J.; Van den Broeck, C. Brownian duet: A novel tale of thermodynamic efficiency. {\em Phys. Rev. X} {\bf 2016}, {\em 6}, 041010.
\bibitem{ef8}Vroylandt, H.; Bonfils, A.; Verley, G. Efficiency fluctuations of small machines with unknown losses. {\em Phys. Rev. E} {\bf 2016}, {\em 93}, 052123.
\bibitem{ef9} Park, J.-M.; Chun, H.-M.;  Noh, J. D. Efficiency at maximum power and efficiency fluctuations in a linear Brownian heat-engine model. {\em Phys. Rev. E}  {\bf 2016}, {\em 94}, 012127.
\bibitem{ef10}Proesmans, K.; Van den Broeck, C. The underdamped Brownian duet and stochastic linear irreversible thermodynamics. {\em Chaos} {\bf 2017}, {\em 27}, 104601.
\bibitem{ef11} Manikandan, S. K.; Dabelow, L.; Eichhorn, R.; Krishnamurthy, S. Efficiency fluctuations in microscopic machines. {\em Phys. Rev. Lett.} {\bf 2019}, {\em122}, 140601.



\bibitem{ergotropy} Allahverdyan, A. E.; Balian, R.;  Nieuwenhuizen, Th. M. Maximal work extraction from finite quantum systems. {\em EPL }{\bf 2004}, {\em 67}, 565.
\bibitem{esposito-mukamel-RMP} Esposito, M.; Harbola, U.; Mukamel, S. Nonequilibrium fluctuations, fluctuation theorems, and counting statistics in quantum systems. {\em Rev. Mod. Phys. }{\bf 2009}, {\em 81}, 1665. 
\bibitem{Barra2015}  Barra, F. The thermodynamic cost of driving quantum systems by their boundaries. {\em Sci. Rep.}  {\bf 2015}, {\em 5}, 14873. 
\bibitem{Chiara} De Chiara, G.; Landi, G. T.; Hewgill, A.; Reid, B.; Ferraro, A.; Rocanglia, A.J.; Antezza, M.  Reconciliation of quantum local master equations with thermodynamics. {\em New J. Phys. }{\bf 2018}, {\em 20} 113024.
\bibitem{esposito-NJP} Esposito, M.; Lindenberg, K.; Van den Broeck, C. Entropy production as correlation between system and reservoir. {\em New J. Phys.} {\bf 2010}, {\em 12} 013013.

\bibitem{StochPRE} Barra, F.; Lled\'o, C. Stochastic thermodynamics of quantum maps with and without equilibrium. {\em Phys. Rev. E} {\bf 2017}, {\em 96}, 052114.  
\bibitem{StochPRE2} Barra, F.; Lled\'o, C. The smallest absorption refrigerator: the thermodynamics of a system with quantum local detailed balance. {\em Eur. Phys. J. Spec. Top.}  {\bf 2018}, {\em 227}, 231. 

\bibitem{terry2} Lostaglio, M.; Korzekwa, K.; Jennings, D.; Rudolph, T. Quantum coherence, time-translation symmetry, and thermodynamics. {\em Phys. Rev. X} {\bf 2015} , {\em 5}, 021001. 
\bibitem{terry3} Lostaglio, M.; Jennings, D.; Rudolph, T. Description of quantum coherence in thermodynamic processes requires constraints beyond free energy. {\em Nat. Commun. } {\bf 2015}, {\em 6}, 6383.


\bibitem{passivity1}  Pusz, W.;  Woronowicz, S. L. Passive states and KMS states for general quantum systems. {\em Commun. Math. Phys.}  {\bf 1978}, {\em 58}, 273.
\bibitem{passivity2} Lenard, A. Thermodynamical proof of the Gibbs formula for elementary quantum systems. {\em J. Stat. Phys.} {\bf 1978}, {\em 19}, 575.

\bibitem{Feller} Feller, W., {\em An introduction to probability theory and its applications vol.1}; John Wiley \& sons Inc: New York, NY, USA, 1968.  

\end{thebibliography}
\end{document}